\documentclass[aps,twocolumn,showpacs,amsmath,amssymb,superscriptaddress,longbibliography,prx]{revtex4-2}

\usepackage[colorlinks=true,allcolors=blue,hypertexnames=false]{hyperref}
\usepackage{amsmath}
\usepackage{dsfont}
\usepackage{hyperref}
\usepackage{textcomp}
\usepackage{tabularx}
\usepackage{graphicx}
\usepackage{soul}
\usepackage{feynmp-auto}
\usepackage{footmisc}
\usepackage[export]{adjustbox}
\usepackage{wrapfig}

\newcommand{\bk}{{\bf k}}
\newcommand{\bx}{{\bf x}}
\newcommand{\bq}{{\bf q}}

\newcommand{\bb}{{\bf b}}

\newcommand{\bc}{{\bf c}}

\newcommand{\bd}{{\bf d}}
\newcommand{\bg}{{\bf g}}

\newcommand{\bF}{{\bf F}}

\def\bH{{\bf H}}
\def\bh{{\bf h}}

\newcommand{\bp}{{\bf p}}

\newcommand{\ve}{{\varepsilon}}

\newcommand{\beg}{\begin{equation}}
\newcommand{\en}{\end{equation}}

\newcommand \bel  {\begin{align}}
\newcommand \enl  {\end{align}}

\newcommand{\veps}{\varepsilon}

\newcommand{\up}{\uparrow}
\newcommand{\dn}{\downarrow}
\newcommand{\dg}{^\dagger}

\def \be{\begin{equation}}
\def \ee{\end{equation}}
\def \bea{\begin{eqnarray}}
\def \eea{\end{eqnarray}}

\begin{document}

\title{Nonunitary spin-triplet superconductors in Zeeman magnetic field}

\author{Wen Li}
\affiliation{Department of Physics, Carnegie Mellon University, Pittsburgh, Pennsylvania 15213, USA}

\author{Vahid Hassanzade}
\affiliation{Department of Physics, Carnegie Mellon University, Pittsburgh, Pennsylvania 15213, USA}

\author{Maxim Dzero}
\affiliation{Department of Physics, Kent State University, Kent, Ohio 44242, USA}

\author{Vladyslav Kozii}
\affiliation{Department of Physics, Carnegie Mellon University, Pittsburgh, Pennsylvania 15213, USA}

\begin{abstract}
We study spin-triplet superconductivity with both unitary and nonunitary pairing in the presence of an external Zeeman magnetic field. Within a mean-field framework, we exactly diagonalize the Bogoliubov–de Gennes Hamiltonian and derive general expressions for the quasiparticle spectrum, superconducting gap, critical temperature, and spin magnetization, valid for arbitrary magnetic-field strengths and temperatures. We analyze in detail the nonlinear spin susceptibility and the field evolution of the superconducting gap and transition temperature, highlighting qualitative differences between unitary and nonunitary pairing states. Our results are broadly applicable to a wide range of materials, including systems with both weak and strong spin–orbit coupling. We show that systematic measurements of the critical temperature and spin susceptibility as functions of the magnitude and orientation of the magnetic field provide a powerful means to identify the structure of the spin-triplet order parameter, and we discuss implications of our findings for candidate materials such as 4Hb-TaS$_2$ and PrOs$_4$Sb$_{12}$.
\end{abstract}

\maketitle

\section{Introduction}

Strong interest in spin-triplet superconductivity is closely tied to its deep connection with topological superconductivity~\cite{SatoAndo2017,AndoFu2015}. This interest has been driven in large part by the growing number of experimental candidates, most prominently uranium-based compounds such as UTe$_2$ \cite{Ran2019,Ran2019NP,Ran2020,Jiao2020,Sakai2022,Ajeesh2023,Theuss2024,Li2025}, UGe$_2$ \cite{Saxena2000,Huxley2001}, U$_{1-x}$Th$_x$Be$_{13}$~\cite{UThBe13review}, UCoGe~\cite{Huy2007,Deguchi2010,Hattori2014Review,Manago2019}, URhGe \cite{Aoki2001}, and UPt$_3$~\cite{UPt3nonunitaryPRL98,nonunitaryUPt3JPSJ96,UPt3review2002}. Spin-triplet pairing has also been proposed in a diverse set of materials, including the Pr-based skutterudite PrOs$_4$Sb$_{12}$ \cite{PrSpinTriplet1,PrSpinTripletMaki,AokiPr,PrOs4Sb12PRL2005,PrOs4Sb12PRL2006,PrOs4Sb12PRB2009}, the van der Waals superconductor 4Hb-TaS$_2$ \cite{4Hb-TaS2_2020,silber2024two}, rhombohedral graphene~\cite{zhou2022isospin,spintripletsuperconductivityrhombohedral}, doped topological insulator Bi$_2$Se$_3$ exhibiting nematic superconductivity~\cite{Zhengetal2016,Maenoetal2017,Willaetal2018,nematicvortexexp2018,Fu2014,nematic2016,Hc22016}, and organic superconductors such as $\lambda$-(BETS)$_2$FeCl$_4$ \cite{Clark12}.

A key experimental hallmark of spin-triplet superconductivity is its unusual robustness against external magnetic fields. Beyond the well-known cases where superconductivity is induced by a magnetic field~\cite{Meul1984,Uji2001}, many candidate materials display a remarkably high tolerance to magnetic fields applied along specific crystallographic directions or coexist with intrinsic ferromagnetic order. This behavior stands in sharp contrast to conventional spin-singlet superconductors, where orbital and Zeeman pair-breaking effects lead to rapid suppression of the superconducting state~\cite{GorkovRusinov,AbrikosovBook}.

Within the family of spin-triplet superconductors, states with {\it nonunitary} pairing constitute a distinctive subclass in which Cooper pairs carry a finite spin on the Fermi surface, giving rise to nondegenerate, spin-selective Bogoliubov quasiparticles~\cite{Leggett75,SigristUeda1991,WolfleBook,volovik1992exotic,He3BlackHole}. It was theoretically predicted that, when combined with strong spin–orbit coupling, nonunitary pairing can realize a new class of ``Majorana superconductors'': topological phases hosting protected chiral nodes analogous to those in Weyl semimetals~\cite{Majorana2016,nematic2016}. The resulting low-energy excitations are bulk three-dimensional Majorana fermions, complementing the surface- and vortex-core–localized Majorana modes anticipated in other spin-triplet superconductors. Several materials have been proposed as candidates for Majorana superconductivity, including UPt$_3$~\cite{UPt3nonunitaryPRL98,nonunitaryUPt3JPSJ96,UPt3review2002}, PrOs$_4$Sb$_{12}$, and PrPt$_4$Ge$_{12}$~\cite{GumeniuketalPRL2008,PrPt4Ge12JPSJ2010,PrPt4Ge12PRB2010,PrPt4Ge12PRB2014,Majorana2016}. Nevertheless, compelling experimental evidence for Majorana superconductivity — and for nonunitary pairing more broadly — remains elusive.

Until recently, nonunitary superconductors received relatively little attention. This was due in part to their extreme rarity, even among the already unconventional class of unitary spin-triplet superconductors, and in part to the greater theoretical complexity required to describe such phases compared to their unitary counterparts. In recent years, however, nonunitary superconductivity has attracted growing interest, motivated by new experimental observations — most notably in the superconducting phases of UTe$_2$. These developments call for a comprehensive theoretical framework for nonunitary superconducting states, as well as for clear and unambiguous experimental signatures enabling their identification.

A recent work by Bernat, Meyer, and Houzet partially addressed this problem~\cite{nonunitarySC2023}. Using a quasiclassical approach~\cite{Larkin1965,LarkinVertex,Eilenberger1968,BELZIG1999}, they generalized the textbook expression for the zero-field spin susceptibility to nonunitary superconductors. This framework, however, has important limitations. First, it is not easily extendable beyond the linear response with respect to magnetic field. Second, it assumes a constant normal-state density of states at the Fermi level, thereby neglecting particle–hole asymmetry effects from the outset. As we show below, this approximation omits several nontrivial features that are particularly relevant for nonunitary superconductors. More broadly, much of the theoretical literature on spin-triplet superconductivity remains confined to the linear-response regime, leaving a range of experimentally relevant situations insufficiently explored.

In this work, we address these issues by developing a general theory of spin-triplet superconductivity in the presence of a Zeeman magnetic field, applicable to a broad class of materials. We diagonalize the Bogoliubov–de Gennes (BdG) Hamiltonian for centrosymmetric spin-triplet superconductors at arbitrary Zeeman field strength and temperature, provided these remain below the ultraviolet cutoff of the theory (typically set by the Debye frequency), and treat both unitary and nonunitary pairing on equal footing. This formulation enables one, in principle, to calculate {\it any observable} associated with the spin response in an external magnetic field.

We first derive the most general expression for the Bogoliubov quasiparticle spectrum, which determines thermodynamic quantities such as the specific heat. We then illustrate our approach by computing the spin susceptibility at finite magnetic field. Finally, we derive the gap equation and analyze how the superconducting transition temperature depends on the magnitude and orientation of the field, discussing how such measurements can be used to identify spin-triplet pairing. Although our analysis is based on a mean-field framework, several of our results extend beyond conventional BCS theory. In particular, we show that particle–hole asymmetry near the Fermi level in the normal state can enhance nonunitary superconductivity at low fields, leading to an increased transition temperature and a finite magnetization even in zero field. These unique effects provide clear and experimentally accessible signatures of nonunitary superconductivity.

Our results are particularly relevant in regimes where orbital effects of the magnetic field can be neglected, such as superconducting thin films subjected to parallel fields and heavy-fermion compounds. They apply to two- and three-dimensional systems, with either weak or strong spin-orbit coupling. Moreover, our framework can be straightforwardly extended to other classes of systems, including multiband and noncentrosymmetric superconductors. From a broader perspective, this work provides a unified description of physical phenomena arising from the coupling between electron spin and an external magnetic field in superconductors, and offers practical tools for the unambiguous experimental identification of spin-triplet pairing, and nonunitary states in particular.

The remainder of the paper is organized as follows. In Sec.~\ref{Sec:Formalism}, we introduce the BdG Hamiltonian in a Zeeman magnetic field and derive the spectrum of Bogoliubov quasiparticles. In Sec.~\ref{Sec:Gapeqn}, we obtain the most general gap equation, valid in both the weak and strong spin–orbit coupling regimes, and analyze its implications for the critical temperature and critical field in Sec.~\ref{SectionTc}. Section~\ref{Sec:magnetization} is devoted to the derivation and discussion of the spin magnetization and spin susceptibility. Applications to systems with specific crystal symmetries relevant to candidate materials are presented in Sec.~\ref{Applications}. We conclude with a discussion and outlook in Sec.~\ref{Sec:discussion}. Technical details of the BdG Hamiltonian diagonalization and the derivation of the transition temperature are provided in the Appendices, along with complementary treatments of spin-singlet pairing and the Green’s-function formalism. Throughout this paper, we use the units with $\hbar = k_B = 1$.


\section{Formalism and single-particle spectrum} \label{Sec:Formalism}
We consider a crystal with inversion and time-reversal symmetry, which ensures that Bloch states at each momentum are double degenerate and form Kramers doublet. We assume that a single double-degenerate band crosses the Fermi level, so the normal-state band Hamiltonian with the Zeeman coupling between electrons and magnetic field $\bH$ has form
\begin{equation}\label{EQ1}
\begin{aligned}
\hat{\cal H}_0=\sum\limits_{\bk\alpha\beta}\hat{c}_{\bk\alpha}^\dagger\left(\xi_{\bk}\delta_{\alpha\beta}-\bh_\bk\cdot{\mbox{\boldmath $\sigma$}}_{\alpha\beta}\right) \hat{c}_{\bk\beta}.
\end{aligned}
\end{equation}
Here, $\xi_\bk$ is the single-particle energy spectrum measured relative to the chemical potential, ${\mbox{\boldmath $\sigma$}}$ is the vector of Pauli matrices, $\hat{c}_{\bk\alpha}^\dagger$ ($\hat{c}_{\bk\alpha}$) are the fermionic creation (annihilation) operators, $\bk$ is crystal momentum and $\alpha,\beta = \uparrow,\downarrow$ is the Kramers degeneracy index. In systems without SO coupling, this index coincides with the spin projection on a preferred axis. Even in the presence of SO coupling, where spin is not a good quantum number, one can still in most cases introduce a ``pseudospin'' index that transforms as spin-1/2 at the $\Gamma$ point. We will refer to it as ``spin'' hereafter for simplicity. For possible exceptions and a comprehensive discussion of this matter, see Ref.~\cite{Samokhin2019}. Here we also define ``momentum-dependent'' magnetic field
\be  \label{Eq:hk}
\bh_\bk = \frac{\mu_B}2 \hat g_\bk \bH,
\ee 
where $\mu_B$ is the Bohr magneton and the real-valued matrix $\hat g_\bk=\hat g_{-\bk}$ describes a possibly momentum-dependent Landé factor. While this matrix becomes diagonal and the Landé factor is close to 2 in the absence of SO coupling, $[\hat{g}_\bk]_{ij} \approx 2 \delta_{ij}$, it may take a rather nontrivial form in crystals with strong SO coupling, leading to significant modifications in, e.g., the spin susceptibility~\cite{Samokhin2021}. 

To study spin-triplet superconductivity, we introduce the mean-field pairing Hamiltonian  $\hat{\cal H}_\Delta$~\cite{SigristUeda1991,MineevSamokhinbook}:
\beg\label{HDelta}
\hat{\cal H}_\Delta=\frac{1}{2}\sum\limits_{\bk}\sum\limits_{\alpha \beta}\left\{\Delta_{\alpha\beta}(\bk)\hat{c}_{\bk\alpha}^\dagger\hat{c}_{-\bk\beta}^\dagger+
\Delta_{\alpha\beta}^\dagger(\bk)\hat{c}_{-\bk\alpha}\hat{c}_{\bk\beta}\right\}.
\en
Here $\Delta_{\alpha\beta}(\bk)$ defines the zero-momentum spin-triplet order parameter
\beg\label{TripletOP}
\Delta_{\alpha\beta}(\bk)=i\Delta\left({\mathbf d}_\bk\cdot{\mbox{\boldmath $\sigma$}}{\sigma}^y\right)_{\alpha\beta}, \qquad \langle |\bd_\bk|^2 \rangle_{\hat \bk}=1.
\en
The momentum dependence of the vector ${\mathbf d}_\bk = -\bd_{-\bk}$ is determined by the irreducible representation of the point symmetry group of the underlying crystal structure. Symbol $\langle\ldots\rangle_{\hat \bk}$ means averaging over the directions of $\bk$ on the Fermi surface, and $\hat \bk = \bk / k$.  We will assume throughout this paper that $\Delta$ is a real positive number for definiteness. The possibility of a state with constant $\Delta$ becoming a metastable one at high enough magnetic fields will be discussed separately.

Our first task is to diagonalize the quadratic Hamiltonian $\hat{\cal H} = \hat{\cal H}_0 + \hat{\cal H}_\Delta$ using the Bogoliubov transformation and find its spectrum. To that end, we introduce the four-component Nambu space according to 
\be  \label{Eq:cNambu}
\hat \bc_\bk = \left(\hat c_{\bk \uparrow}, \hat c_{\bk \downarrow}, \hat c^\dagger_{-\bk \uparrow}, \hat c^\dagger_{-\bk \downarrow} \right)^T.
\ee 
In this basis, the mean-field Hamiltonian can be written as 
\be  \label{Eq:HBdG}
\hat{\cal H} = \hat{\cal H}_0 + \hat{\cal H}_\Delta =  \sum_{\bk}\frac12\hat \bc^\dagger_\bk H_{\text{BdG}}(\bk) \hat \bc_\bk + \xi_\bk, 
\ee 
where the Bogoliubov – de Gennes Hamiltonian is given by 
\be  \label{Hbdg}
\qquad H_{\text{BdG}}(\bk) = \left(  \begin{array}{cc}  \xi_{\bk} - \bh_\bk \cdot {\boldsymbol \sigma} & \Delta_{\bk} \\ \Delta_{\bk}^\dagger & -\xi_{\bk} + \bh_\bk \cdot{\boldsymbol \sigma}^T     \end{array} \right).
\ee 
Here we used the identities $\xi_\bk = \xi_{-\bk}$ and $\bh_\bk = \bh_{-\bk}$ guaranteed by inversion and time-reversal symmetries of the normal state. The diagonalization of this Hamiltonian is straightforward but rather cumbersome, so we present the technical details and the explicit form of the unitary Bogoliubov transformation in Appendix~\ref{App:Diagonalization}. We find for the spectrum of the Bogoliubov quasiparticles: 

\be\label{Eks}
E_{\bk, \pm}= \sqrt{\xi_\bk^2 + \Delta^2|\bd_\bk|^2 + h_{\bk}^2\pm r_\bk}, 
\ee
where we introduced for brevity
\be\label{rk}
r_\bk=\sqrt{\left(\Delta^2 \bq_\bk - 2\xi_\bk \bh_\bk\right)^2+ 4 \Delta^2 |{\mathbf d}_\bk \cdot {\bh_\bk}|^2}, \quad h_\bk = |\bh_\bk|.  
\ee
The real vector $\bq_\bk = i \bd_\bk \times \bd_\bk^*$ is nonzero only in nonunitary superconducting states. Physically, it is proportional to the total spin of a Cooper pair on the Fermi surface, ${\bf S}_\bk \propto \bq_\bk$~\cite{Leggett75,MineevSamokhinbook}. In the unitary case, defined by $\Delta(\bk)\Delta^{\dagger}(\bk) \propto I$, where $I$ is the identity matrix, this vector vanishes, and we recover the expressions obtained in Refs.~\cite{PangZhou2025,Meyer2024}.

Equations~\eqref{Eks}-\eqref{rk} is the first important result of our work. It can be used directly to determine the low-temperature behavior of the specific heat for various states, and at various fields. 
We stress that this result, as well as other results obtained in this work, can be independently derived using the formalism of the Green's functions, as we explicitly demonstrate in Appendix~\ref{App:Gorkov}.


\section{Gap equation} \label{Sec:Gapeqn}
To determine the dependence of the order parameter on the temperature and magnetic field, we derive the corresponding gap equation. To that end, we take a step back and write down the interaction Hamiltonian in the Cooper channel:
\be  \label{Eq:HI}
H_I = \frac12\sum_{\bk,\bp} V_{\alpha \beta, \lambda\mu}(\bk,\bp) \hat{c}^\dagger_{-\bk\alpha}\hat{c}^\dagger_{\bk\beta}\hat{c}_{\bp\lambda}\hat{c}_{-\bp\mu}. 
\ee 
Here and below, we assume a summation over repeated (pseudo-)spin indices $\alpha$, $\beta$, $\lambda$, $\mu$. The explicit form of the interaction potential $V_{\alpha \beta, \lambda\mu}(\bk,\bp)$ depends on the crystal symmetry and the presence or absence of spin-orbit coupling, as we discuss in the following.

The superconducting order parameter within the mean-field approximation is defined as follows: 
\beg\label{MainSC}
\begin{aligned}
\Delta_{\bp,\alpha\beta}&=\sum\limits_{\bk}V_{\beta\alpha,\lambda\mu}(\bp,\bk)\langle\hat{c}_{-\bk\mu}\hat{c}_{\bk\lambda}\rangle.
\end{aligned}
\en
The symbol $\langle \ldots\rangle$ here should be understood as quantum mechanical and thermal averaging. Using the explicit form of the Bogoliubov transformation from Appendix~\ref{App:Diagonalization}, we find for this anomalous average:
\be  \label{Eq:g}
\langle\hat{c}_{-\bk\mu}\hat{c}_{\bk\lambda}\rangle = - \bF_\bk \cdot\bg_{\lambda \mu}, \quad \text{with} \qquad \bg_{\lambda \mu}= \left( i {\boldsymbol \sigma} \sigma^y \right)_{\lambda \mu}
\ee 
and 
\be \label{Eq:Fk}
\begin{aligned}  
\bF_\bk = &\Delta\sum_{s= \pm}\frac{\tanh(E_{\bk, s}/2T)}{4E_{\bk, s}r_\bk}\left[r_\bk\, \bd_\bk + 2s(\bd_\bk\cdot{\bh_\bk}) \bh_\bk\right.\\&\left.+2is\,\xi_\bk \bd_\bk \times \bh_\bk -i s\Delta^2\bd_\bk \times \bq_\bk\right].
\end{aligned}
\ee

Equations~\eqref{MainSC}-\eqref{Eq:Fk}, along with definition~\eqref{TripletOP}, allow us, in principle, to determine the structure and value of the order parameter for the most generic pairing interaction, assuming that the gap function is spin-triplet. To derive a more explicit and practically applicable form of the gap equation, we do a number of approximations and focus on the specific cases of strong and negligible SO coupling.

In crystals with {\it strong spin-orbit coupling}, the pairing potential can be written as follows~\cite{MineevSamokhinbook}:
\be\label{Interaction}
\begin{aligned}
V_{\alpha\beta,\lambda\mu}(\bk,\bp)=\frac{1}{2}\sum_{\Gamma}V_\Gamma(k,p) \\ \times\sum\limits_{i=1}^{d_\Gamma}\left({\mbox{\boldmath $\psi$}}_i^\Gamma(\hat{\bk}) \cdot {\mathbf g}_{\alpha\beta}\right)\left({\mbox{\boldmath $\psi$}}_{i}^{\Gamma*}(\hat{\bp}) \cdot {\mathbf g}\dg_{\lambda\mu}\right),
\end{aligned}
\ee
where ${\mbox{\boldmath $\psi$}}_i^\Gamma(\hat{\bk})$ are the vector basis  functions of a certain odd-parity irreducible representation $\Gamma$ of the crystal point symmetry group, with dimensionality $d_\Gamma$. Vector of matrices ${\mathbf g}_{\alpha\beta}$ is defined in Eq.~\eqref{Eq:g}, and the symbol ${\mathbf g}\dg_{\lambda\mu}$ should be understood as $({\mathbf g}\dg)_{\lambda\mu}$. In the spirit of the BCS theory, we also assume that
\beg\label{VGammakkp}
V_\Gamma(k,p)=\left\{\begin{array}{cc} -v_\Gamma, & -\varepsilon_0^\Gamma\leq\xi_\bk,\xi_{\bp}\leq \varepsilon_0^\Gamma, \\
0, & \mathrm{otherwise.}
\end{array}
\right.
\en
The attraction in a given representation implies $v_\Gamma > 0$, and the energy scale $\varepsilon_0^\Gamma > 0$ sets the ultraviolet cutoff, which is typically of the order of Debye frequency $\omega_D$. 

To simplify our analysis further, we only keep a single representation in Eq.~\eqref{VGammakkp} with the highest transition temperature and most robust superconductivity, and suppress the index $\Gamma$ in basis functions for brevity~\footnote{This approximation may fail deep in the superconducting phase or if different channels are nearly degenerate, such that the basis functions from different representations can mix.}. The $\bd_\bk$-vector can now be expanded in the basis of the eigenfunctions of this representation: 
\beg\label{dtopsi}
{\mathbf d}_\bk=\sum\limits_{i=1}^{d_\Gamma}\eta_i{\mbox{\boldmath $\psi$}}_i(\hat{\bk}), \qquad \sum\limits_{i=1}^{d_\Gamma}|\eta_i|^2=1.
\en
Complex coefficients $\eta_i$, when multiplied by $\Delta$, should be viewed as a multi-component order parameter that enters, e.g., the Ginzburg-Landau free energy expansion. These coefficients must be chosen to minimize the free energy and generically depend on the microscopic details. Plugging in this expansion and Eq.~\eqref{Interaction} into the gap equation~\eqref{MainSC} and using orthonormality of the basis functions, 
\be 
\langle {\mbox{\boldmath $\psi$}}_i(\hat{\bk}) \cdot {\mbox{\boldmath $\psi$}}_j^*(\hat{\bk})\rangle_{\hat \bk} = \delta_{ij},
\ee 
we find that $\Delta$ and $\eta_i$ must satisfy the set of equations
\be  \label{Eq:gapeqn1}
\Delta \cdot \eta_i = v_\Gamma \sum_{\bk} {\mbox{\boldmath $\psi$}}_i^*(\hat{\bk})\cdot \bF_\bk.
\ee 
We stress that these equations must be satisfied for each $i=1,\ldots,d_\Gamma$. 

To derive a single equation for $\Delta$, assuming that $\eta_i$ are fixed, we can use the normalization condition in Eq.~\eqref{dtopsi} and find that
\be  \label{Eq:gapeqgeneric}
\Delta = v_{\Gamma} \sum_{\bk} \bd_{\bk}^* \cdot \bF_{\bk}, 
\ee 
which, given definition~\eqref{Eq:Fk}, finally leads to 
\begin{equation}\label{SelfEq}
\begin{aligned}
\frac{1}{v_\Gamma} = \sum_{\bk, s}&\frac{\tanh{(E_{\bk, s}/2T)}}{4E_{\bk, s}r_\bk}\left(r_\bk\, |\bd_\bk|^2+2s|\bd_\bk\cdot\bh_\bk|^2\right.\\&\left.-2s\,\xi_\bk\bq_\bk\cdot\bh_\bk+s\Delta^2 q_\bk^2\right),
\end{aligned}
\end{equation}
where the summation is over the Bogoliubov band index $s=\pm$. Equation~\eqref{SelfEq} is the second important result of our work, as it allows us to calculate the value of the superconducting gap at any values of temperature and Zeeman field. In Sec.~\ref{Applications} below, we demonstrate how $\Delta$ evolves with temperature and magnetic field for different gap structures.

It is straightforward to show that Eq.~\eqref{SelfEq} also holds in crystals with {\it negligible spin-orbit} coupling. Indeed, in this case, the spin rotation symmetry dictates the following form of the pairing interaction~\cite{MineevSamokhinbook}: 
\be\label{InteractionnoSO}
\begin{aligned}
&V_{\alpha\beta,\lambda\mu}(\bk,\bp)=\frac{1}{2}\bg_{\alpha \beta} \cdot \bg^\dagger_{\lambda \mu} \\ &\times\sum_{\Gamma}V_\Gamma(k,p)\sum\limits_{i=1}^{d_\Gamma}\psi_i^\Gamma(\hat{\bk}) \psi_{i}^{\Gamma*}(\hat{\bp}),
\end{aligned}
\ee
where $\psi_i^\Gamma(\hat \bk)$ are now the scalar basis functions of the irreducible representation $\Gamma$ satisfying orthonormality condition. Focusing again on a single representation with the most robust superconductivity, the $\bd_\bk$-vector can be parametrized as
\beg \label{Eq:detanoSO}
{\mathbf d}_\bk=\sum\limits_{i=1}^{d_\Gamma}{\boldsymbol \eta}_i\psi_i(\hat{\bk}), \qquad \sum\limits_{i=1}^{d_\Gamma}|{\boldsymbol \eta}_i|^2=1,
\en
where the components of the order parameter ${\boldsymbol \eta}_i$ are now vectors. The set of gap equations in this case becomes
\be  
\Delta \cdot {\boldsymbol \eta}_i = v_\Gamma \sum_{\bk} \psi_i^*(\hat{\bk})\cdot \bF_\bk.
\ee 
Multiplying this equation by ${\boldsymbol \eta}^*_i$, summing over $i$, and using  the normalization condition in Eq.~\eqref{Eq:detanoSO}, one ends up again with Eqs.~\eqref{Eq:gapeqgeneric} and~\eqref{SelfEq}.

Finally, we show that the expression for the energy of the superconducting condensate has the same form as without the magnetic field. Starting from the general formula for the spin-triplet condensate
\be  
{\cal E}_0 = \Delta\sum_{\bk} \bd_\bk \cdot \bF^*_\bk + \frac{1}2\sum_{\bk,s}\left(\xi_{\bk,s} - E_{\bk,s}\right),
\ee 
where $\xi_{\bk,+} = \xi_{\bk,-} = \xi_{\bk}$ due to Kramers degeneracy, and using Eq.~\eqref{Eq:gapeqgeneric}, we end up with the familiar expression
\be\label{E0Familiar}
{\cal E}_0 = \frac{\Delta^2}{v_\Gamma} + \frac{1}2\sum_{\bk,s}\left(\xi_{\bk,s} - E_{\bk,s}\right).
\ee 
We stress that although this expression does not explicitly depend on the magnetic field, it does so implicitly through $\Delta$ and $E_{\bk,s}$. This becomes more evident if we replace the $1/v_\Gamma$ factor using Eq.~\eqref{SelfEq}. 

The free energy of a superconductor is then given by 
\begin{align}  \label{Eq:F}
F_S &= {\cal E}_0 - T \sum_{\bk, s} \ln \left(1 + e^{-E_{\bk,s}/T}\right)\\ &= \frac{\Delta^2}{v_\Gamma} + \frac{1}{2}\sum_{\bk, s} \left[{\xi_{\bk,s}} - 2T\ln\left(2 \cosh\frac{E_{\bk,s}}{2T}  \right)\right], \nonumber
\end{align} 
which coincides with the expression for the thermodynamic potential $\Omega$ \cite{Svidzinski}.

\section{Magnetic field dependence of the critical temperature}\label{SectionTc}
Another useful probe that can be directly extracted from the gap equation, except for $\Delta$ itself, is the magnetic field dependence of the transition temperature $T_c$. 

Assuming the transition is second-order, we set $\Delta\to 0$ in Eq.~\eqref{SelfEq}  and find the following equation for the critical temperature: 
\be\label{Tc}
\begin{aligned}
\ln \left(\frac{T_c}{T_{c0}}\right) = &  \left\langle \frac{|\bd_\bk \cdot \bh_\bk|^2}{h_{\bk}^2} \cdot F\left( \frac{h_\bk}{2T_c}  \right)\right\rangle_{\hat{\bk}} \\+ \frac{\nu_F'}{\nu_F} &\left\langle \left(\bq_\bk \cdot{\mathbf \bh_\bk}\right) \ln\left( \frac{\omega_D}{\max\{ h_\bk, T_c  \}} \right)\right\rangle_{\hat \bk}, \\ 
F(x) = \text{Re}&\left[\psi\left(\frac12\right) - \psi\left(\frac12+\frac{ix}{\pi}\right)\right].
\end{aligned}
\ee
Here, $\psi(x)$ is the digamma function,  $T_{c0}$ is the transition temperature at zero field, $\nu_F$ is the normal-state density of states at Fermi energy per one spin projection, and $\nu_F'$ is its derivative. We also use the Debye frequency $\omega_D$ as an ultraviolet cutoff of the theory, i.e., take $\ve_0^\Gamma = \omega_D$ in Eq.~\eqref{VGammakkp}, which implies that expression~\eqref{Tc} is valid provided $\max\{h_\bk, T_c\}\ll \omega_D$. Finally, when switching from the summation over momenta $\bk$ to integration over $\xi$, we assume a spherical Fermi surface for simplicity. The technical details of this derivation are discussed in Appendix~\ref{App:Tc}. In the limit of a constant density of states, $\nu_F'=0$, we reproduce the answer from Ref.~\cite{FrigerietallPRL2004}.

To simplify the analysis, we assume for the rest of this section that the matrix of Landé factors is isotropic and momentum-independent, $[\hat{g}_\bk]_{ij} \approx 2 \delta_{ij}$, such that the Zeeman magnetic field is given by $\bh_\bk \approx \mu_B \bH$.
We consider the case of unitary pairing first, $\bq_\bk = 0$, such that vector $\bd_\bk$ can be chosen real-valued, up to a possible overall phase factor. It is convenient then to rewrite Eq.~\eqref{Tc} in the form
\be\label{Analytics}
\ln \left(\frac{T_c}{T_{c0}}\right) = \zeta \cdot F\left( \frac{\mu_B H}{2T_c} \right),  ~ \zeta = \left\langle{|\bd_\bk \cdot\hat \bH|^2}\right\rangle_{\hat \bk},
\ee 
where $\hat \bH={\mathbf H}/H$ is a unit vector along ${\mathbf H}$.
If $\bd_\bk \perp \bH$ for all $\bk$, then $\zeta=0$ and the magnetic field does not affect $T_c$ at all, $T_c(H) = T_{c0}$. Furthermore, at large enough fields $\mu_B H \gg T_{c0}$ and as long as $\zeta < 1$ we find that~\cite{Sigrist2009}
\be {\label{Eq:Tcasymptotic}}
T_c \approx T_{c0} \left( \frac{\pi T_{c0}}{2 e^\gamma \mu_B H} \right)^{\zeta/(1-\zeta)},
\ee 
 where $\gamma\approx 0.577$ is the Euler's constant. To derive this expression, we used the identity $\psi(1/2) = -\gamma - \ln 4$.
 
\begin{figure}
\includegraphics[width=0.90\linewidth]{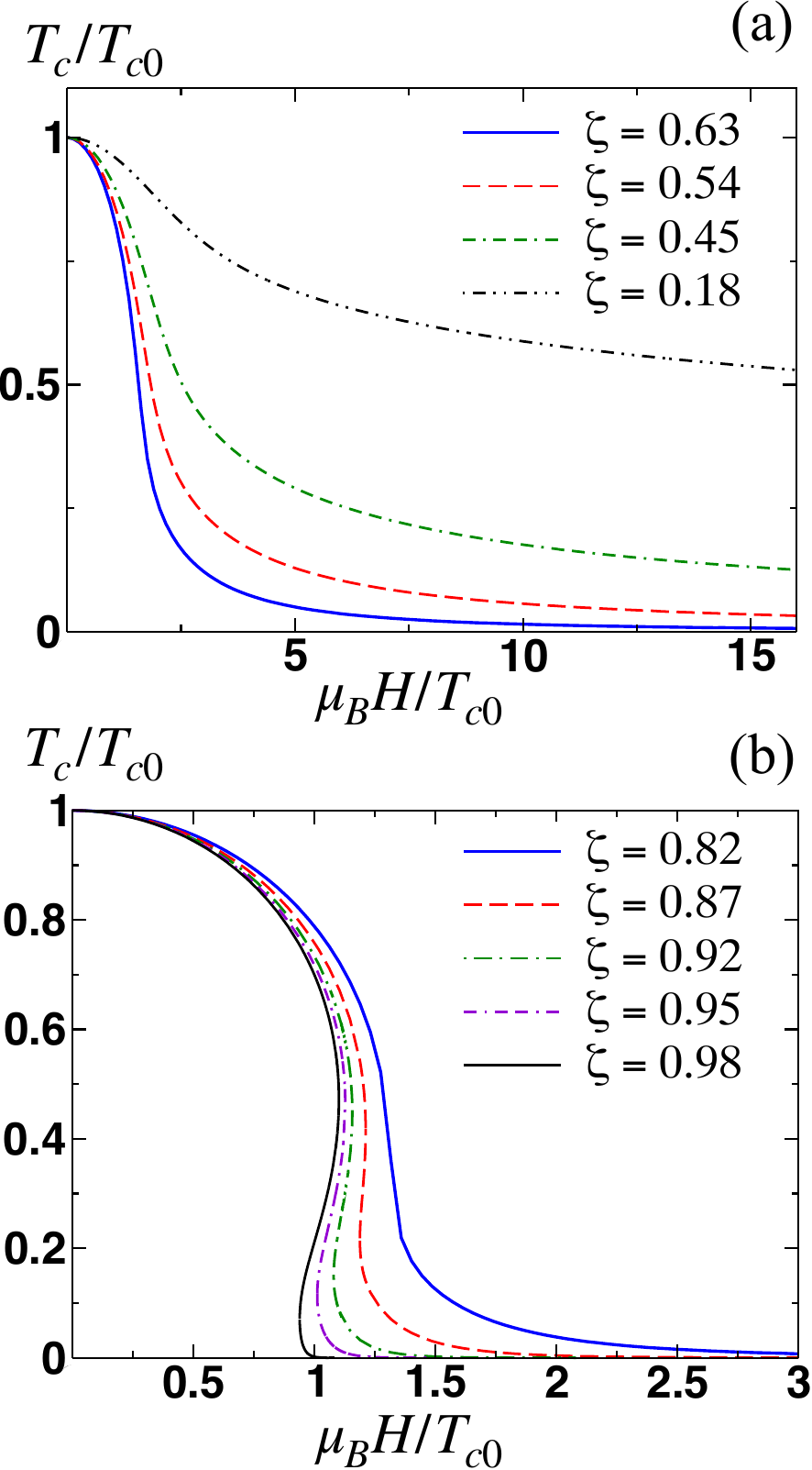} 
\caption{The dependence of the superconducting critical temperature $T_c$ on the magnitude of the Zeeman magnetic field $H$ for the case of a unitary spin-triplet superconductor, $\bq_\bk=0$. All the parameters are measured in the units of the critical temperature at zero magnetic field $T_{c0}$. (a) The curves are obtained by numerically evaluating Eq.~\eqref{SelfEq} for the gap function in $E_u$ representation of the group $O_h$, see Sec.~\ref{Applications}. The results are in perfect agreement with Eq.~\eqref{Analytics}. (b) Numerical evaluation of Eq.~\eqref{Analytics} at $\zeta\ge0.82$. Multiple solutions for $T_c$ at a given $H$ indicate the metastable phase and the first-order transition.} 
\label{Fig2-Unitary-Tc}
\end{figure}

Figure~\ref{Fig2-Unitary-Tc} shows the characteristic dependence of the critical temperature on the magnetic field for different values of $\zeta$. This behavior suggests that such measurements offer a powerful probe of the symmetry of the superconducting order parameter. Since different field orientations correspond to different values of $\zeta$ (see Sec.~\ref{Applications} and Fig.~\ref{Fig-Zeta}), fitting experimental data at multiple angles allows one, in principle, to infer the form of the $\bd_\bk$-vector from Eq.~\eqref{Analytics}.

Figure~\ref{Fig2-Unitary-Tc}(b) shows that for $\zeta > 0.83$ the solution for $T_c$ becomes multivalued at certain fields, signaling the emergence of a metastable state and hence a first-order phase transition. To determine the transition line, one must compare the free energy of the superconducting state — given by Eq.~\eqref{Eq:F} with $\Delta$ satisfying the gap equation~\eqref{SelfEq} — to that of the normal state with $\Delta = 0$. This analysis was carried out in detail in Ref.~\cite{Meyer2024}.

The nonmonotonic behavior of $T_c(H)$ in Fig.~\ref{Fig2-Unitary-Tc}(b) is common in systems where superconductivity competes with magnetism or other pair-breaking interactions. Similar profiles have been reported in the context of excitonic instabilities in narrow-gap semiconductors~\cite{Barzykin2000}, magnetic-field–induced gapless superconductivity in multiband systems~\cite{Barzykin2009}, and conventional superconductivity in Kondo lattices~\cite{MHZ1970,MHZ1971,Dzero2005,Fisher2005,Barzykin2005,Awelewa2024,Awelewa2025}. In the latter case, the dependence of $T_c$ on the concentration of Kondo centers is controlled by the ratio $T_c/T_K$, where $T_K$ is the single-ion Kondo temperature. When $T_c \ll T_K$, the impurity moment becomes fully screened, effectively rendering the impurity nonmagnetic; the Anderson theorem then ensures that $T_c$ remains finite, albeit small.  In the other examples, however, the characteristic multivalued dependence of $T_c$ on magnetic field points to an instability toward a spatially inhomogeneous state accompanied with the first-order transition.

In the case of nonunitary pairing, $\bq_\bk \neq 0$, the right-hand side of Eq.~\eqref{Tc} acquires an additional term which, for $\bh_\bk = \mu_B \bH$, becomes
$$\frac{\nu_F'}{\nu_F} \mu_B \bH\cdot \left\langle \bq_\bk \right\rangle_{\hat \bk} \ln\left( \frac{\omega_D}{\max\{ \mu_B H, T_c  \}} \right).$$
If $\langle\bq_\bk\rangle_{\hat{\bk}} = 0$, the equation for $T_c$ reduces to the unitary form and depends only on the parameter $\zeta$, Eq.~\eqref{Analytics}, even though the $\bd_\bk$-vector is complex. By contrast, when $\langle\bq_\bk\rangle_{\hat{\bk}} \neq 0$, this extra term introduces strong sensitivity to the direction of the magnetic field. In particular, if $\langle\bq_\bk\rangle_{\hat{\bk}}$ and $\bH$ are parallel for $\nu_F' > 0$ (or antiparallel for $\nu_F' < 0$), the transition temperature actually {\it increases} with field, $dT_c/dH > 0$, at sufficiently small $\mu_B H \ll T_{c0}$. This feature can be viewed as an indication of nonunitary superconductivity. Within the BCS theory, this contribution is often neglected, since $\nu_F' \neq 0$ requires particle–hole asymmetry and is typically small at large Fermi energy. However, it may become appreciable in low-density superconductors or near Van Hove singularities.

Finally, we note that Eqs.~\eqref{Tc}–\eqref{Eq:Tcasymptotic} can be inverted and interpreted as describing the critical field as a function of temperature, $H_c(T)$, assuming a second-order transition. Regions with $dH_c/dT > 0$ signal metastability and a first-order transition, as discussed above. Moreover, Eq.~\eqref{Eq:Tcasymptotic} shows that the Pauli limiting field diverges as $T \to 0$ for $\zeta < 1$, implying that orbital effects must then be included~\cite{FrigerietallPRL2004,Meyer2024}.

\section{Spin Magnetization} \label{Sec:magnetization}
We now turn our attention to magnetization induced by Zeeman magnetic field. 
Given definitions~\eqref{EQ1} and~\eqref{Eq:hk}, spin magnetization of a superconductor can be calculated as~\cite{Leggett75,MineevSamokhinbook,Samokhin2021}
\be\label{DefineM}
{\mathbf M}=\frac{\mu_B}2\sum\limits_{\bk}\hat g_\bk^T\langle \hat{c}_{\bk\alpha}^\dagger{\mbox{\boldmath $\sigma$}}_{\alpha\beta}\hat{c}_{\bk\beta}\rangle.
\ee
Symbol $\langle \ldots \rangle$ here means quantum mechanical and thermal average. Straightforward evaluation using Bogoliubov transformation from Appendix~\ref{App:Diagonalization} yields~\footnote{In the limit $\Delta\to 0$, we have $E_{\bk, s}\to|\xi_\bk+s h_\bk|$ and recover the familiar expression for the magnetization of the ideal Fermi gas ${\mathbf M}_{n}\propto\sum_{\bk, s}s\tanh(|\xi_\bk+s h_\bk|/2T){\mathbf h}_\bk$.}
\begin{align}\label{MExplicit}
{\mathbf M} &=\frac{\mu_B}2\sum_{\bk,s=\pm} \hat g_\bk^T\frac{ \tanh(E_{\bk, s}/2T)}{2E_{\bk, s}r_\bk}\left\{\big(r_\bk + 2s\xi_\bk^2\big)\bh_\bk\right.\\&\left.+s\Delta^2 ({\mathbf d}_\bk^*\cdot\bh_\bk){\mathbf d}_\bk +s\Delta^2({\mathbf d}_\bk\cdot\bh_\bk){\mathbf d}_\bk^*-s\xi_\bk\Delta^2 {\mathbf q}_\bk\right\}. \nonumber
\end{align}

Note that the last term in the brackets in Eq.~\eqref{MExplicit} does not explicitly depend on magnetic field. As such, it defines
the {\it field-independent} part of spin magnetization: 
\be \label{Eq:M0general} 
{\bf M}_0 \approx \frac{\mu_B}2\nu_F' \Delta^2 \ln\left(\frac{\omega_D}{\Delta}\right) \left\langle \hat g_\bk^T \bq_\bk \right\rangle_{\hat \bk}, 
\ee 
where $\nu_F'$, as before, is the derivative of the single-particle density of states at the Fermi energy. In the case of isotropic and momentum-independent Landé factor, $[\hat{g}_\bk]_{ij} = 2 \delta_{ij}$, Eq.~\eqref{Eq:M0general} simplifies to 
\be \label{Eq:M0}
{\bf M}_0 \approx \mu_B\nu_F' \Delta^2 \ln\left(\frac{\omega_D}{\Delta}\right) \left\langle \bq_\bk \right\rangle_{\hat \bk}.  \tag{\ref*{Eq:M0general}$^{\prime}$}
\ee 
We note that $\Delta$, in principle, is a function of the magnetic field $\bH$. This expression has similar origin and form as the second term in Eq.~\eqref{Tc}.

It follows from Eq.~\eqref{Eq:M0general} that the zero-field spin magnetization is a unique feature of nonunitary superconductivity (provided the normal state is not magnetic), and not just any pairing that breaks time-reversal symmetry. 
We also stress that it is not solely determined by the average spin of the Cooper pairs $\left\langle \bq_\bk \right\rangle_{\hat \bk}$, even if the latter is nonzero. It also requires particle-hole asymmetry near the Fermi energy in the normal state, $\nu_F' \ne 0$. This term is quite small within the BCS theory, as typically $\nu_F' \propto 1/\ve_F$. However, we expect that it may become important and potentially measurable in low-density superconductors or if the Fermi energy resides sufficiently close to Van Hove singularities, providing a clear indication of the nonunitary spin-triplet pairing. In addition, the particle-hole asymmetry can also be induced by subjecting a superconductor to an external terahertz electromagnetic light, and under this condition this term may play a role in determining the magnetic response. 


\subsection{Zero-field spin susceptibility}
To extract the zero-field spin susceptibility, $\chi_{ij}^{(0)} = \left(\partial M_i/\partial H_j   \right)\vert_{H\to 0}$, we use Eqs.~\eqref{MExplicit} and~\eqref{Eq:hk} and assume constant density of states $\nu_F$ for simplicity, which amounts to neglecting the contributions $O(\Delta/\veps_F)$:
    \be \label{Eq:susc}
    \begin{aligned}
&\chi_{ij}^{(0)}  = \frac{\chi_n}4\left\langle [\hat{g}_\bk^T]_{i l}\left\{ \mathcal{N}(\bk, T)\left[\delta_{lm} - \mathrm{Re}(\hat d_{\bk l} \hat d^*_{\bk m})\right] \right.\right.\\&\left.\left. + \mathcal{Y}(\bk, T)\mathrm{Re}(\hat d_{\bk l} \hat d^*_{\bk m})   +\mathcal{Q}(\bk, T) \hat q_{\bk l} \hat q_{\bk m}\right\} [\hat{g}_\bk]_{m j}\right\rangle_{\hat \bk},
\end{aligned}
    \ee 
where $\hat d_{\bk i} = d_{\bk i} /|\bd_\bk|$, $\chi_n = 2\mu_B^2 \nu_F$ is the spin susceptibility of a normal metal, and the summation over repeated indices $l$ and $m$ is implied. Here we have introduced the dimensionless functions 
\begin{align}
&\mathcal{N}(\bk, T) = \int_{0}^{\infty} d\xi\sum_{s = \pm}\frac{ \tanh(\ve_{s}/2T)}{2\ve_{s}}\left(1 + \frac{2 s \xi^2}{\Delta^2 q_\bk}  \right), \nonumber  \\ 
&\mathcal{Y}(\bk, T) = \int_{0}^{\infty} d\xi\sum_{s = \pm} \frac{\left(2s\ve^2_s -\Delta^2 q_\bk\right) \tanh(\ve_{s}/2T)}{
2\Delta^2 q_\bk\ve_{s}}, \nonumber \\ 
&\mathcal{Q}(\bk, T) = \int_{0}^{\infty} d\xi\sum_{s = \pm} \frac{ \xi^2}{4 \Delta^2 q_\bk} \left( \frac{\Delta^2 q_\bk}{T \ve_s^2 \cosh^2(\ve_s/2T)} \right.  \nonumber \\ &\left.- \frac{2 \tanh(\ve_s/2T)(\Delta^2 q_\bk+2s \ve_s^2)}{\ve_s^3}   \right),
\end{align}
where $\ve_{\pm} = (\xi^2 + \Delta^2 |{\mathbf d}_\bk|^2 \pm \Delta^2 q_\bk)^{1/2}$. In the unitary limit, $\bq_\bk \to 0$ (which also implies that $\mathcal{Q} \to 0$), we reproduce the expression obtained in Ref.~\cite{Samokhin2021}.

Focusing again on the isotropic $\bk$-independent Landé factor, $[\hat{g}_\bk]_{ij} = 2 \delta_{ij}$, one can rewrite Eq.~\eqref{Eq:susc} as 
   \be 
    \begin{aligned}
\chi_{ij}^{(0)} &= {\chi_n}\left\langle \left\{ \mathcal{N}(\bk, T)\left[\delta_{ij} - \mathrm{Re}(\hat d_{\bk i} \hat d^*_{\bk j})\right] \right.\right.\\&\left.\left. + \mathcal{Y}(\bk, T)\mathrm{Re}(\hat d_{\bk i} \hat d^*_{\bk j})   +\mathcal{Q}(\bk, T) \hat q_{\bk i} \hat q_{\bk j}\right\}\right\rangle_{\hat \bk}.
\end{aligned} \tag{\ref*{Eq:susc}$^{\prime}$}
    \ee 
This expression reproduces the one derived in Ref.~\cite{nonunitarySC2023}. It is straightforward to show that 
in the unitary limit, ${\mathbf q_\bk}\to 0$, we reproduce the textbook expressions~\cite{Leggett75,MineevSamokhinbook} 
\beg\label{Standartchi}
\mathcal{N}(\bk, T)\to 1, ~\mathcal{Y}(\bk, T) \to Y(\bk, T), ~\mathcal{Q}(\bk, T)\to0, 
\en
where 
\be\label{Eq:Yoshida}
Y(\bk, T) = \frac1{2T}\int_0^{\infty}\frac{d\xi}{\cosh^2(\sqrt{\xi^2 + \Delta^2 |\bd_\bk|^2 }/2T)}
\ee
is the standard Yoshida function. Similarly, at $T=T_c$, we find that 
\be 
\mathcal{N}(\bk, T_c) = \mathcal{Y}(\bk, T_c) = 1, \qquad \mathcal{Q}(\bk, T_c)=0,
\ee
{leading to the susceptibility of a normal metal.}
Finally at $T= 0$, we obtain 
\begin{align}
&\mathcal{N}(\bk, 0) = \frac12\left(1 + \frac{|\bd_\bk|^2}{2q_\bk}\ln\frac{|\bd_\bk|^2 + q_\bk}{|\bd_\bk|^2-q_\bk} \right),  \\ &\mathcal{Y}(\bk, 0) = \mathcal{Q}(\bk, 0) = \frac12\left(1 - \frac{|\bd_\bk|^2}{2q_\bk}\ln\frac{|\bd_\bk|^2 + q_\bk}{|\bd_\bk|^2-q_\bk} \right). \nonumber
\end{align}
These expressions have singularity in the extreme nonunitary case, i.e., when $q_\bk = |\bd_\bk|^2$ (equivalently, $\bd^2_\bk = 0$). One can show that in this case the susceptibility at $T=0$ coincides exactly with the unitary expression ($\bq_\bk = 0$): 
\be 
\chi_{ij}\left(T=0, q_\bk = |\bd_\bk|^2\right) = \chi_n \left\langle \left[\delta_{ij} - \mathrm{Re}(\hat d_{\bk i} \hat d^*_{\bk j}) \right] \right\rangle_{\hat \bk}.
\ee

All results in this subsection, obtained for the case $[\hat{g}_\bk]_{ij} = 2 \delta_{ij}$, are consistent with the expressions reported in Ref.~\cite{nonunitarySC2023}, even though they appear in a different form. The difference stems from the fact that Ref.~\cite{nonunitarySC2023} employs the quasiclassical Eilenberger formalism, where one first integrates over the electronic dispersion $\xi$ (assuming a constant density of states), and all final expressions are written as sums over Matsubara frequencies. While this approach correctly reproduces the linear spin susceptibility, it makes generalizing the results to the nonlinear regime — or deriving the constant part of the magnetization, Eq.~\eqref{Eq:M0} — considerably more challenging.

\subsection{Nonlinear spin susceptibility: qualitative discussion}
General expression for magnetization, Eq.~\eqref{MExplicit}, allows us to evaluate the nonlinear spin susceptibility $\chi_{ij}({\mathbf H}) = \partial M_i/\partial H_j$ at arbitrary value of the magnetic field $H$ (assuming it is below the ultraviolet cutoff energy $\omega_D$). Even though the closed analytical expression for $\chi_{i j}({\mathbf H})$ can be written down, it turns out to be quite cumbersome, so we find it more constructive to calculate $\chi_{i j}({\mathbf H})$ numerically and discuss main qualitative features specific for various types of pairing. In the calculation of the susceptibility it is important to keep in mind that the order parameter $\Delta$ itself bears dependence on the magnetic field, which should be found self-consistently from the gap equation~\eqref{SelfEq}.

The most striking feature of the nonlinear susceptibility is the sharp peak that appears precisely where the order parameter $\Delta$ varies rapidly with magnetic field. When the field orientation produces a sufficiently large pair-breaking parameter $\zeta$, $\Delta$ is abruptly suppressed within a narrow field interval, giving rise to a pronounced peak in the nonlinear response, as illustrated by the examples in the next section. As the field direction is rotated and $\zeta$ decreases, $\Delta(H)$ evolves more smoothly, causing the peak to broaden, diminish to a small bump, and eventually disappear, corresponding to the regime in which the order parameter becomes essentially field independent. This peak, originating from the rapid suppression of $\Delta$, is a generic feature of both unitary and nonunitary pairing states.

\section{Application to various systems}\label{Applications}

We now apply our results to systems with various lattice symmetries, assuming strong spin–orbit coupling and an odd-parity (spin-triplet) order parameter. For simplicity, we take $g_{ij}(\bk) = 2 \delta_{ij}$ throughout this section and assume that the magnetic field does not alter the pairing symmetry. We also briefly comment on candidate spin-triplet superconductors compatible with these symmetry settings, while emphasizing that concrete applications to real materials are significantly more subtle and require additional analysis, which we leave for future work.

\subsection{Cubic group $O_h$ }
We start with the cubic point group $O_h$ which applies to the heavy-fermion compounds U$_{1-x}$Th$_x$Be$_{13}$~\cite{UThBe13review} and PrOs$_4$Sb$_{12}$~\cite{PrSpinTriplet1,PrSpinTripletMaki,AokiPr,PrOs4Sb12PRL2005,PrOs4Sb12PRL2006,PrOs4Sb12PRB2009}, the latter of which was suggested as a candidate for a Majorana superconductor~\cite{Majorana2016}. It has three multidimensional odd-parity irreducible representations (irreps): $E_u$, $T_{1u}$, and $T_{2u}$. Here we focus on the two-dimensional irrep $E_u$ for definiteness. Its basis functions to the leading $p$-wave order are given by 
\beg\label{BasisEuOh}
{\begin{aligned}
{\mbox{\boldmath $\psi$}}_1({\hat{\bk}})&=\frac{1}{\sqrt{2}}\left(-\hat{k}_x\hat {\bx}-\hat{k}_y\hat {\bf y}+2\hat{k}_z\hat {\bf z}\right), \\
{\mbox{\boldmath $\psi$}}_2({\hat{\bk}})&=\sqrt{\frac{3}{2}}\left(\hat{k}_x\hat {\bx}-\hat{k}_y\hat {\bf y}\right),
\end{aligned}}
\en
such that the $\bd$-vector is defined as $\bd_\bk=\sum\limits_{a=1}^2\eta_a{\mbox{\boldmath $\psi$}}_a({\hat{\bk}})$.

\begin{figure}
\includegraphics[width=0.90\linewidth]{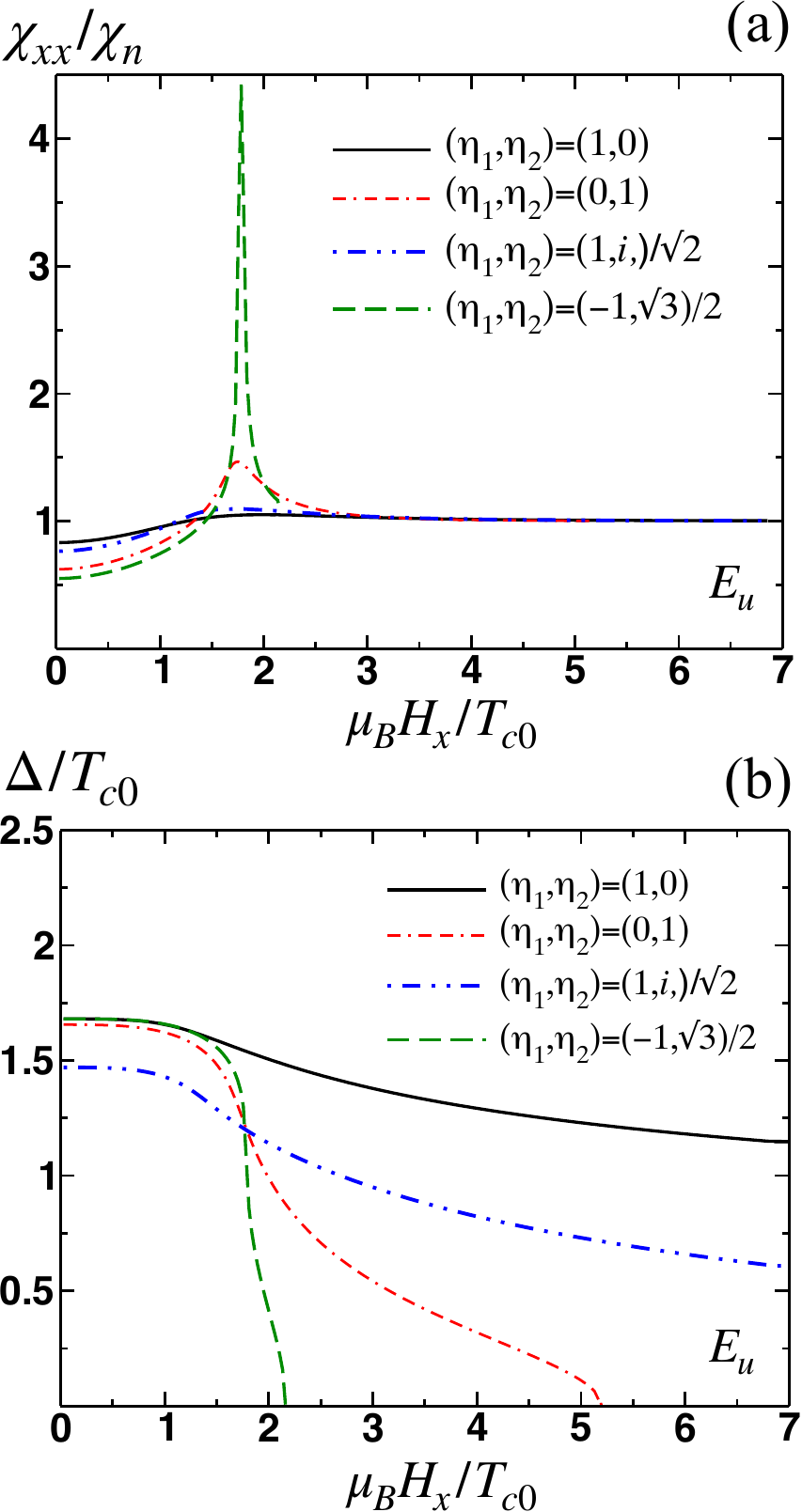} 
\caption{$E_u$ representation of the cubic point group $O_h$. (a)~The dependence of the nonlinear spin susceptibility $\chi_{xx}$, measured in units of the normal-state value $\chi_n$, on the magnitude of the Zeeman magnetic field $H_x$, for different choices of the order parameter $(\eta_1,\eta_2)$. The figure is numerically evaluated by differentiating Eq.~\eqref{MExplicit}. (b)~The dependence of the superconducting gap $\Delta$ on the Zeeman field $H_x$. The figure is numerically evaluated from the gap equation~\eqref{SelfEq}. The peak in $\chi_{xx}$ appears exactly at fields where $\Delta$ decreases abruptly. $\Delta$ and $\mu_B H_x$ are measured in units of the zero-field transition temperature $T_{c0}$. The temperature is taken to be $T=T_{\mathrm{c0}}/6$.} 
\label{Fig:Eu}
\end{figure}

Depending on the form of the order parameter $(\eta_1,\eta_2)$, different “flavors” of multicomponent superconductivity can be realized. When the order parameter is real, the system enters a unitary nematic phase that breaks the crystal’s rotational symmetry~\cite{nematic2016,Hc22016,nematicCDW2019}. In this case, two symmetry-inequivalent states (not related by rotational operations) are possible: $(\eta_1,\eta_2) = (1,0)$ and $(\eta_1,\eta_2) = (0,1)$, corresponding to fully gapped and nodal superconducting states, respectively.

In the opposite scenario, a complex order parameter $(\eta_1,\eta_2) = (1,i)/\sqrt{2}$ breaks time-reversal symmetry and realizes a chiral Majorana superconductor~\cite{Majorana2016,nematic2016}. In this case, the $\bd$-vector takes the form $\bd_\bk = (\ve \hat k_x, \ve^2 \hat k_y, \hat k_z)$, with $\ve = e^{2\pi i/3}$, leading to
\be 
\bq_\bk = \sqrt{3} (\hat k_y \hat k_z, \hat k_x \hat k_z,\hat k_x \hat k_y).
\ee 
As a consequence, one pseudospin species of Bogoliubov quasiparticles exhibits point nodes along directions satisfying $\hat k_x^2 = \hat k_y^2 = \hat k_z^2 = 1/3$, giving rise to itinerant three-dimensional Majorana fermions as low-energy excitations, while the second species remains fully gapped. We note that for this gap structure $\langle \bq_\bk \rangle_{\hat\bk} = 0$.

\begin{figure}
\includegraphics[width=0.950\linewidth]{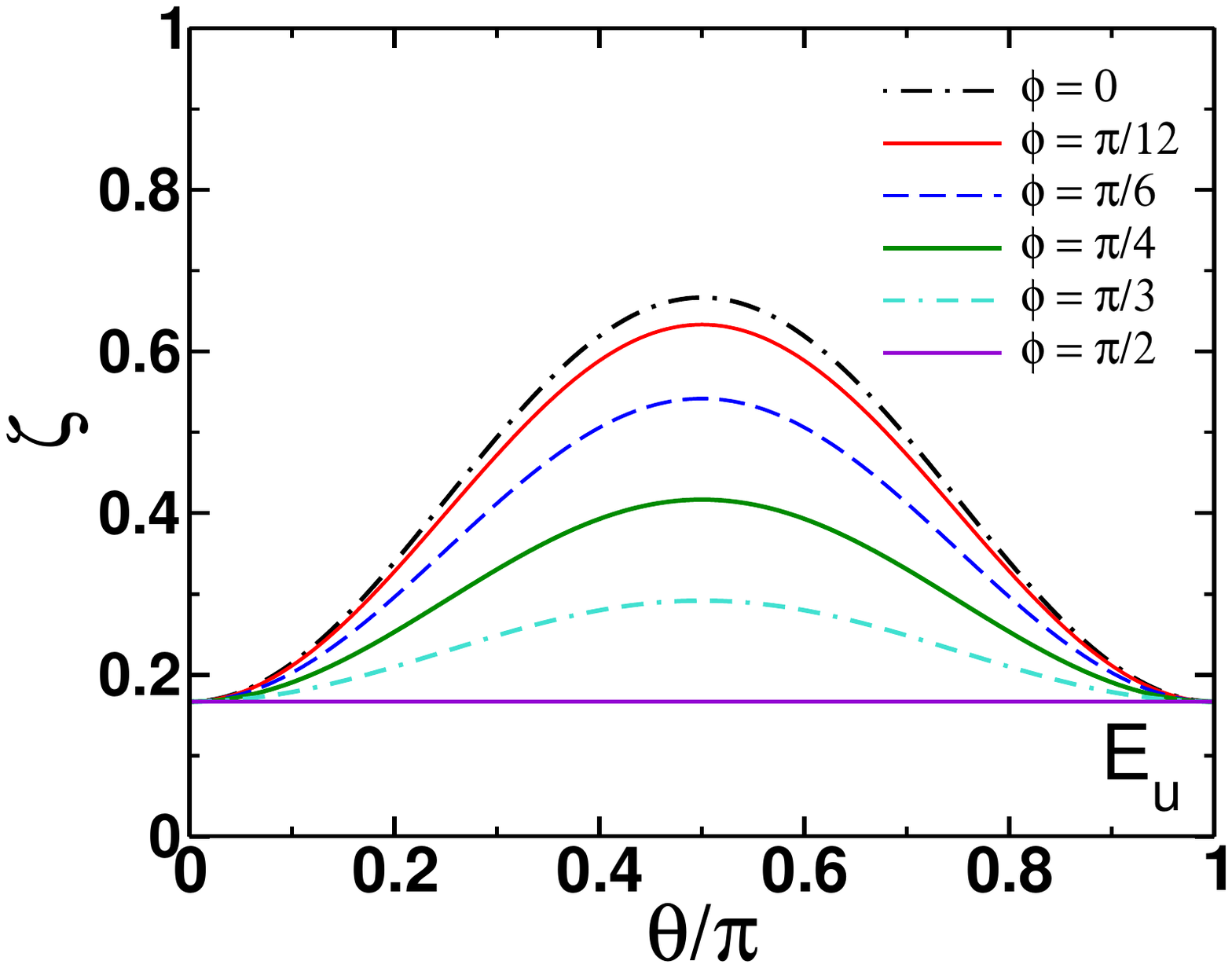} 
\caption{The dependence of parameter $\zeta$, Eq. \eqref{Analytics}, on the magnetic field direction, ${\mathbf H}=H(\cos\phi\sin\theta,\sin\phi\sin\theta,\cos\theta)$. The vector ${\mathbf d}_\bk$ here is determined by the $E_u$ representation of $O_h$ with $(\eta_1,\eta_2) = (-1,\sqrt{3})/2$.} 
\label{Fig-Zeta}
\end{figure}

The evolution of the $\chi_{xx}$ component of the nonlinear spin susceptibility and superconducting gap $\Delta$ as a function of magnetic field, for different choices of $(\eta_1,\eta_2)$, is shown in Fig.~\ref{Fig:Eu}. A pronounced peak appears at field values that also correspond to an abrupt suppression of the superconducting gap $\Delta$. We find this behavior to be universal across different lattice symmetries and representations, for both unitary and nonunitary pairings. We further note that although the order parameter $(\eta_1,\eta_2) = (-1,\sqrt{3})/2$ is symmetry-related to $(\eta_1,\eta_2) = (1,0)$, their responses differ markedly because the magnetic field is fixed along the $\hat{\bx}$ direction. This observation is fully consistent with the fact that the depairing parameter $\zeta$, which controls the field-induced suppression of $\Delta$, is highly sensitive to the field orientation, as illustrated for $(\eta_1,\eta_2) = (-1,\sqrt{3})/2$ in Fig.~\ref{Fig-Zeta}.

\subsection{Hexagonal group $D_{6h}$}
Our second example is the hexagonal point group $D_{6h}$ which is relevant to UPt$_3$~\cite{UPt3nonunitaryPRL98,nonunitaryUPt3JPSJ96,UPt3review2002} and possibly to the van der Waals superconductor 4Hb-TaS$_2$ \cite{4Hb-TaS2_2020,silber2024two}. It has two two-dimensional odd-parity irreps: $E_{1u}$ and $E_{2u}$. We focus on $E_{2u}$, with the basis functions given by
\beg\label{BasisE2uD6h}
{\begin{aligned}
{\mbox{\boldmath $\psi$}}_1({\hat{\bk}})&=\sqrt{\frac{3}{2}}\left(\hat{k}_x\hat {\bx}-\hat{k}_y\hat {\bf y}\right), \\
{\mbox{\boldmath $\psi$}}_2({\hat{\bk}})&=\sqrt{\frac{3}{2}}\left(\hat{k}_y\hat {\bx}+\hat{k}_x\hat {\bf y}\right).
\end{aligned}}
\en

The real-valued order parameters $(\eta_1,\eta_2) = (1,0)$ and $(\eta_1,\eta_2) = (0,1)$ produce a full superconducting gap in the ${\bf x}$–${\bf y}$ plane and degenerate point nodes along the ${\bf z}$ axis. In contrast, the chiral state with order parameter $(\eta_1,\eta_2) = (1,i)/\sqrt{2}$ realizes nonunitary pairing, in which only one spin species develops a superconducting gap while the other remains completely gapless. This phase  is characterized by $\bd_\bk = (\sqrt{3}/2)(\hat k_x + i \hat k_y)(\hat \bx + i \hat {\bf y})$ and $\bq_\bk = (3/2)(\hat k_x^2+\hat k_y^2) \hat {\bf z}$. Such a gap structure may be relevant to the metallic, $T$-linear specific-heat behavior observed in superconducting 4Hb-TaS$2$~\cite{4Hb-TaS2_2020}, although a more detailed analysis is required to substantiate this conjecture. At the same time, experimental evidence for nematic superconductivity in this material appears more consistent with pairing in the $E{1u}$ irrep~\cite{silber2024two}.

\begin{figure}
\includegraphics[width=0.90\linewidth]{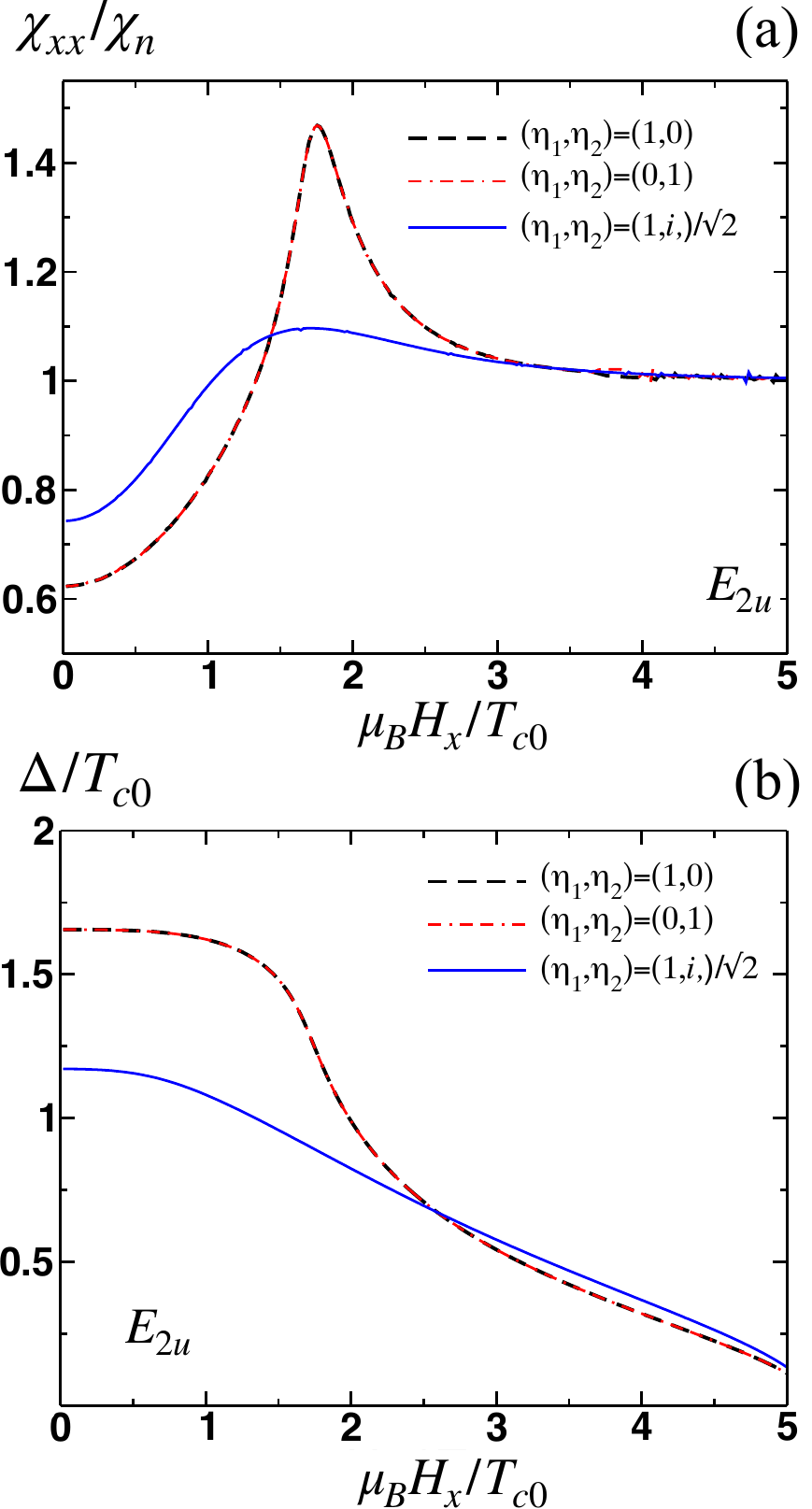} 
\caption{Same as Fig.~\ref{Fig:Eu}, but for the $E_{2u}$ irreducible representation of the $D_{6h}$ point group instead of the $E_u$ representation of $O_h$.} 
\label{Fig:E2u}
\end{figure}

The evolution of $\chi_{xx}$ and $\Delta$ with the Zeeman field is shown in Fig.~\ref{Fig:E2u}. The qualitative behaviour is similar to Fig.~\ref{Fig:Eu}, though the peaks in susceptibility in this case are not so pronounced.

\section{Conclusions}\label{Sec:discussion}
To summarize, we have developed a theory of centrosymmetric spin-triplet superconductors in an external Zeeman magnetic field. By diagonalizing the corresponding BdG Hamiltonian, we derived the Bogoliubov transformation and the quasiparticle spectrum, valid for a broad range of magnetic field strengths and temperatures. We used these results to derive and analyze the gap equation and the nonlinear spin susceptibility. Our framework applies to systems with both weak and strong spin–orbit coupling, in two and three dimensions, and accommodates both unitary and nonunitary pairing. The regime of validity of our theory is determined by the requirement that fluctuations about the mean-field solution remain small, which is typically satisfied when $\max\{\Delta, T, \mu_B H \}\ll\omega_D\ll \veps_F$~\cite{kos2004gaussian,Joerg2018}.

Our results can be directly applied to a broad class of candidate spin-triplet and nonunitary superconductors. We identify several experimentally accessible signatures of these states, including the enhancement of superconductivity by an applied magnetic field and the emergence of a finite magnetization in zero field. We further show that the evolution of the superconducting gap $\Delta$ and the transition temperature $T_c$ in a Zeeman field is largely governed by the dimensionless parameter $\zeta=\left\langle (\bd_\bk\cdot \bH)/H\right\rangle_{\hat \bk}$, which plays the role of an effective pair-breaking parameter and can be tuned experimentally by rotating the magnetic-field direction. We find that superconductivity is only weakly suppressed for $\zeta \ll 1$, whereas the suppression becomes much more pronounced as $\zeta$ approaches unity, potentially leading to metastability and a first-order transition. Moreover, we show that magnetic-field regimes associated with the strongest suppression of $\Delta$ correspond to sharp peaks in the nonlinear spin susceptibility. Taken together, these results provide powerful experimental tools for identifying the superconducting gap structure.

The expressions obtained from diagonalizing the Bogoliubov–de Gennes Hamiltonian, or equivalently from the Green’s-function formalism, can be readily used to compute additional observables, including the nuclear spin–lattice relaxation rate and the magnetic penetration depth. Moreover, our framework can be extended to other classes of superconductors, such as multiorbital and noncentrosymmetric systems~\cite{GorkovRashbaPRL2001}. In the latter case, it is particularly interesting to explore the emergence of finite-momentum Fulde–Ferrell–Larkin–Ovchinnikov states and their behavior in a Zeeman magnetic field~\cite{BarzykinGorkovPRL2002}.

Another promising direction is the study of collective modes in spin-triplet superconductors, especially in the presence of nonunitary, spin-nondegenerate pairing, and the role played by the pair-breaking parameter $\zeta$. In spin-singlet superconductors, similar pair-breaking effects were recently shown to have profound consequences for collective-mode dynamics~\cite{Yantao2024,Li2025,Kamenev2025}. Investigating these phenomena for the spin-triplet superconductors is an equally promising and important direction for the future work.

\section{Acknowledgments}
This work was financially supported by the National Science Foundation grant DMR-2400484 (M.D.). M. D. has performed this work in part at Aspen Center for Physics, which is supported by the National Science Foundation grant PHY-2210452.

\begin{appendix}

\section{Diagonalization of the BdG Hamiltonian for the spin-triplet pairing \label{App:Diagonalization}}

To diagonalize the BdG Hamiltonian~\eqref{Hbdg}, we introduce a unitary Bogoliubov transformation $\hat \bc_\bk = \hat U_\bk \hat \bb_\bk$, where $\bb_{\bk,\pm}$ are the Bogoliubov quasiparticles' annihilation operators satisfying standard anticommutation relations $\left\{\hat \bb_{\bk,s},\hat \bb^\dagger_{\bk,s'}\right\} = \delta_{s,s'},$ $\left\{\hat \bb_{\bk,s}^\dagger,\hat \bb^\dagger_{\bk,s'}\right\} = \left\{\hat \bb_{\bk,s},\hat \bb_{\bk,s'}\right\} = 0$, with $s,s'=\pm$. The particle-hole symmetry dictates that the unitary transformation matrix $\hat U_\bk$ has the form 
\be \label{Eq:U}
\hat U_{\bk}=\left( \begin{array}{cc} u_{\bk} & v_{\bk} \\ v_{-\bk}^* & u_{-\bk}^*  \end{array}    \right), \qquad \hat U_\bk^\dagger \hat U_\bk =I,
\ee 
where $u_{\bk}$ and $v_{\bk}$ are $2\times2$ matrices in the spin space and $I$ is the identity $4\times4$ matrix. The diagonalization can be viewed as a matrix equation for the components of $u_\bk$ and $v_\bk$:
\be
\hat U^\dagger_\bk H_{\text{BdG}}(\bk) \hat U_\bk = \left( \begin{array}{cccc} E_{\bk,+} &0&0&0 \\0 & E_{\bk,-} & 0 &0 \\ 0 & 0& -E_{-\bk,+} & 0 \\ 0 & 0 & 0 & -E_{-\bk,-}    \end{array} \right),
\ee
while the inversion symmetry dictates that $E_{-\bk,\pm} = 
E_{\bk,\pm}$. Eigenenergies $E_{\bk,\pm}$ are given by Eq.~\eqref{Eks}. The solution of this equation is not unique and can be found by different methods. For instance, we show in Appendix~\ref{App:Gorkov} how this diagonalization can be performed using the method of Gor'kov Green's functions. Instead, here we use the approach similar to that proposed in Ref.~\cite{SigristUeda1991} and find a solution in the form  
\begin{align} 
u_\bk&=\sum_{s = \pm}X_{\bk,s} \left( \alpha^0_{\bk,s} \sigma^0 + {\boldsymbol \alpha_{\bk,s}} \cdot {\boldsymbol \sigma}  \right)(\sigma^0+s\sigma^z),  \\ v_\bk&=\sum_{s = \pm}X_{\bk,s} \left( \beta^{0*}_{\bk,s} \sigma^0 + {\boldsymbol \beta^*_{\bk,s}} \cdot {\boldsymbol \sigma}  \right) \sigma^y(\sigma^0+s\sigma^z), \nonumber
\end{align}
where $\sigma^0$ and $\boldsymbol \sigma$ are the identity and the vector of Pauli matrices in spin space. Here we introduced functions
\begin{widetext}
\be 
\begin{aligned}
\alpha^0_{\bk,\pm} &= 2\frac{(\bd_\bk \cdot \bh_\bk)^2}{\bd_\bk^2} \pm r_{\bk}, \qquad {\boldsymbol \alpha_{\bk,\pm}}  = \Delta^2 \bq_\bk - 2 \xi_\bk \bh_\bk + 2\frac{(\bd_\bk \cdot \bh_\bk)}{\bd_\bk^2}\left [(\xi_\bk - E_{\bk,\pm})\bd_\bk - i \bd_\bk \times \bh_\bk \right],  \\ \beta^0_{\bk,\pm} &= \frac{i}{\Delta\bd_\bk^2}\left\{ (\bd_\bk \cdot\bh_\bk)\left[(E_{\bk,\pm}-\xi_\bk)^2-h_\bk^2+\Delta^2|\bd_\bk|^2\right]-i\Delta^2\bh_\bk \cdot(\bq_\bk\times\bd_\bk)
    \right\},  \\ {\boldsymbol \beta_{\bk,\pm}} & = \frac{i}{\Delta\bd_\bk^2}\left\{ \bd_\bk\left[\pm r_\bk(E_{\bk,\pm}-\xi_\bk)-2\xi_\bk h_\bk^2+\Delta^2\bq_\bk \cdot\bh_\bk\right] \right.  \\ &\left.+i\bd_\bk\times\bh_\bk\left[\pm r_\bk - 2\xi_\bk(E_{\bk,\pm}-\xi_\bk)\right]-i\Delta^2\bq_\bk\times\bd_\bk(E_{\bk,\pm}-\xi_\bk)+\Delta^2\bq_\bk (\bd_\bk \cdot\bh_\bk)
    \right\},
\end{aligned}
\ee
with $\bd_\bk^2 \equiv \bd_\bk \cdot \bd_\bk$ and $r_\bk$ given by Eq.~\eqref{rk}. The normalization constants $X_{\bk,\pm}$ should be found from the unitarity condition~\eqref{Eq:U}. After straightforward but lengthy calculation, we obtain

\begin{align}\label{Eq:Xpm}
    &X_{\bk,\pm}=\frac{\Delta\left|\bd_\bk^2\right|}{\sqrt{16r_\bk E_{\bk,\pm}}}\bigg\{ 
    \big[ (r_\bk\mp\Delta^2|\bd_\bk|^2)(E_{\bk,\pm}-\xi_\bk)\mp2\xi_\bk h_\bk^2
    \pm 2\Delta^2\bq_\bk \cdot \bh_\bk\big]\big(|\bd_\bk|^2\mp\bq_\bk \cdot \hat{\bf z}\big)+\big[(E_{\bk,\pm}-\xi_\bk)^2-h_\bk^2\big] \times \nonumber
    \\
    &\times\left[|\bd_\bk|^2\bh_\bk \cdot \hat{\bf z}-(\bd^*_\bk \cdot \bh_\bk)(\bd_\bk \cdot \hat{\bf z})-(\bd_\bk \cdot \bh_\bk)(\bd_\bk^* \cdot \hat{\bf z})\mp\bq_\bk \cdot \bh_\bk\big]\pm \Delta^2|\bd_\bk^2|^2\big[(E_{\bk,\pm}-\xi_\bk)\mp\bh_\bk \cdot \hat{\bf z}\right]
     \bigg\}^{-1/2}.
\end{align}

Equations~\eqref{Eq:U}-\eqref{Eq:Xpm} can now be used to calculate various correlation functions. For example, the anomalous average in Eq.~\eqref{Eq:g} can be expressed as
\be
\langle\hat{c}_{-\bk\mu}\hat{c}_{\bk\lambda}\rangle = u_{\bk, \lambda +}v_{\bk, \mu +} (1-f_{\bk,+})- u_{\bk, \mu +} v_{\bk, \lambda +} f_{\bk,+} + u_{\bk, \lambda -}v_{\bk, \mu -} (1-f_{\bk,-})- u_{\bk, \mu -} v_{\bk, \lambda -} f_{\bk,-},
\ee
where $f_{\bk,\pm} = (1+\exp[E_{\bk,\pm}/T])^{-1}$ is the Fermi-Dirac distribution function, and we used that at thermal equilibrium $\langle b^\dagger_{\bk,s} b_{\bk,s'}\rangle = \delta_{s s'} - \langle b_{\bk,s'} b^\dagger_{\bk,s}\rangle = f_{\bk,s} \delta_{ss'}$. Applying the Bogoliubov transformation derived above, we find that
\begin{equation}
\begin{array}{rcr}
u_{\bk,\lambda +} v_{\bk,\mu +} & = & i\frac{\Delta}{2r_\bk}\left\{ \left[(\bd_\bk \cdot \bh_\bk) \sigma^0 - \frac{1}{2E_{\bk,+}} \left( r_\bk \bd_\bk + 2(\bd_\bk\cdot \bh_\bk) \bh_\bk +2 i \xi_\bk \bd_\bk \times \bh_\bk  + i \Delta^2 \bq_\bk\times\bd_\bk  \right) \cdot{\boldsymbol \sigma}    \right] \sigma^y \right\}_{\lambda\mu}, \\
u_{\bk,\lambda -} v_{\bk,\mu -} & = & -i\frac{\Delta}{2r_\bk}\left\{ \left[(\bd_\bk \cdot \bh_\bk) \sigma^0 + \frac{1}{2E_{\bk,-}} \left( r_\bk \bd_\bk - 2(\bd_\bk \cdot\bh_\bk) \bh_\bk -2 i \xi_\bk \bd_\bk \times \bh_\bk - i \Delta^2\bq_\bk\times\bd_\bk \right) \cdot {\boldsymbol \sigma}    \right] \sigma^y \right\}_{\lambda\mu}.
\end{array}
\end{equation}
Collecting everything together, we end up with Eqs.~\eqref{Eq:g}-\eqref{Eq:Fk}. 

Absolutely analogously, after straightforward but cumbersome algebra, we derive Eq.~\eqref{MExplicit} for spin magnetization. 
\end{widetext}

\section{Critical temperature \label{App:Tc}}
To derive the expression for critical temperature (and critical field), we start with Eq.~\eqref{SelfEq} and put $\Delta = 0$, assuming the transition is second-order:
\begin{equation}
\begin{aligned} \label{Eq:AppTc}
&\sum_{\bk,s=\pm}\frac{\tanh{[(|\xi_\bk|+ s h_\bk)/2T_c]}}{4|\xi_\bk| h_\bk (|\xi_\bk| + s h_\bk)}\left(|\xi_\bk| h_\bk |\bd_\bk|^2+s|\bd_\bk\cdot\bh_\bk|^2\right.\\&\left.-s\,\xi_\bk\bq_\bk\cdot\bh_\bk\right) = \sum_{\bk}\frac{\tanh(|\xi_\bk|/2T_{c0})}{2|\xi_\bk|} |\bd_\bk|^2 = \frac{1}{v_\Gamma},
\end{aligned}
\end{equation}
where $T_{c0}$ is the transition temperature at zero field. 
To evaluate this expression, we assume a spherically symmetric Fermi surface and switch from the summation over $\bk$ to integration over $\xi_\bk$ and averaging over the Fermi surface: 
\be
\sum_{\bk}\left\{\ldots\right\} = \nu_F \int d\xi_\bk \langle\ldots\rangle_{\hat \bk}, \qquad \langle\ldots\rangle_{\hat \bk} =  \int \frac{d\Omega_{\hat \bk}}{S_d} \left\{\ldots\right\}.
\ee 
The element of a solid angle is denoted by $d \Omega_{\hat \bk}$, and $S_d$ is the area of a  unit sphere in $d$ dimensions: $S_1 = 2$, $S_2 = 2\pi$, $S_3 = 4\pi$ etc.
To simplify this gap equation further, we extend the upper limit of integration in Eq.~\eqref{Eq:AppTc} to $\infty$, which is justified in the limit $\omega_D \gg \max\{T_c,h_\bk\}$. Using then the identity 
\be  
\int_0^\infty \left( \frac{\tanh(\xi+x)}{\xi+x} + \frac{\tanh(\xi-x)}{\xi-x} - 2\frac{\tanh\xi}\xi \right) d\xi = 0,
\ee 
which is valid for any real number $x$, we can rewrite it as
\begin{align}  \label{Eq:AppTc2}
&\sum_{\bk}\frac{|\bd_\bk \cdot \bh_\bk|^2 - \xi_\bk \bq_\bk \cdot \bh_\bk}{2|\xi_\bk| h_\bk} \sum_{s = \pm}s\frac{\tanh\frac{|\xi_\bk| + s h_\bk}{2T_c}}{|\xi_\bk| + s h_\bk} = \nonumber \\ &\sum_{\bk} |\bd_\bk|^2 \frac{\tanh\frac{|\xi_\bk|}{2T_{c0}}-\tanh\frac{|\xi_\bk|}{2T_{c}}}{|\xi_\bk|} = 2 \nu_F \ln\frac{T_c}{T_{c0}}. 
\end{align}
To evaluate the right-hand side of this equation, we used the identity 
\be  
\int_0^{\infty}\frac{d\xi}{\xi}\left(\tanh\frac{\xi}{2T_{c0}} - \tanh\frac{\xi}{2T_{c}}\right) = \ln \frac{T_c}{T_{c0}},
\ee 
along with the normalization condition $\langle |\bd_\bk|^2\rangle_{\hat \bk} = 1$. To calculate the first term on the left-hand side, we use the equality 
\be  
\int_0^{\infty} \frac{d\xi}\xi \left(\frac{\tanh(\xi+x)}{\xi+x} - \frac{\tanh(\xi-x)}{\xi-x}   \right) = \frac2x F(x),
\ee 
where function $F(x)$ is defined in Eq.~\eqref{Tc}. This gives us 
\begin{align}
\sum_{\bk}&\frac{|\bd_\bk \cdot \bh_\bk|^2 }{2|\xi_\bk| h_\bk} \left(\frac{\tanh\frac{|\xi_\bk| + h_\bk}{2T_c}}{|\xi_\bk| + h_\bk} -\frac{\tanh\frac{|\xi_\bk| - h_\bk}{2T_c}}{|\xi_\bk| - h_\bk} \right)   \nonumber \\ & =2 \nu_F \left\langle  \frac{|\bd_\bk \cdot \bh_\bk|^2}{h_\bk^2} F\left(\frac{h_\bk}{2T_c}\right) \right\rangle_{\hat\bk}.
\end{align}
Finally, the second term on the left-hand side of Eq.~\eqref{Eq:AppTc2} vanishes upon integration over $\xi$ if we assume constant density of states at Fermi level, i.e., normal-state particle-hole symmetry. To pick up the leading nonzero contribution, we expand $\nu_F(\xi) \approx \nu_F + \xi \nu_F' + \ldots$. With this expansion, we find to the leading order
\begin{align}
&\sum_{\bk}-\frac{ \xi_\bk \bq_\bk \cdot \bh_\bk}{2|\xi_\bk| h_\bk} \left(\frac{\tanh\frac{|\xi_\bk| + h_\bk}{2T_c}}{|\xi_\bk| + h_\bk} -\frac{\tanh\frac{|\xi_\bk| - h_\bk}{2T_c}}{|\xi_\bk| - h_\bk} \right)   \nonumber \\ &\approx 2 \nu'_F \left\langle  (\bq_\bk \cdot \bh_\bk) \int\limits^{\sim \omega_D}_{\sim \max\{h_\bk, T_c  \}}\frac{d\xi}{\xi}\right\rangle_{\hat\bk} \\ &\approx 2\nu_F' \left\langle  (\bq_\bk \cdot \bh_\bk) \ln\frac{\omega_D}{\max\{ h_\bk, T_c   \}}\right\rangle_{\hat\bk}, \nonumber
\end{align}
where we took into account that the major (logarithmic) contribution to the integral comes from the region $\xi \gg \max\{ h_\bk, T_c\}$. Collecting all the terms together, we reproduce Eq.~\eqref{Tc}.

\section{Spin-singlet pairing}
\label{App:Diagonalizationsinglet}
To make this study comprehensive and self-sufficient, we also apply our approach to spin-singlet pairing in Zeeman magnetic field. Most of the results reported in this Appendix are not new, but we use them to compare with the spin-triplet case. 

The mean-field BdG Hamiltonian is given by the same expression~\eqref{Eq:HBdG}, but now the order parameter is parametrized as
\be  
\Delta_{\alpha\beta}(\bk)=\Delta \phi_\bk (i\sigma^y_{\alpha \beta}), \qquad \langle |\phi_\bk|^2 \rangle_{\hat \bk}=1,
\ee 
where $\phi_{-\bk} = \phi_{\bk}$ is the complex scalar function. Upon diagonalization, we find that the dispersion of the Bogoliubov quasiparticles is given by  
\be
E_{\bk, \pm}= \ve_\bk\pm h_\bk, \,\,\,\, \ve_\bk = \sqrt{\xi_\bk^2+\Delta^2 |\phi_\bk|^2},
\ee
with $h_\bk = |\bh_\bk|$. The unitary transformation matrix $\hat U_\bk$ has the same structure as in Eq.~\eqref{Eq:U}, but now its components are give by
\begin{align} 
u_\bk&= X_{\bk}  \sum_{s = \pm} \left( \bh_\bk \cdot {\boldsymbol \sigma} - s h_{\bk} \sigma^0 \right)(\sigma^0+s\sigma^z),  \\ v_\bk&=iX_{\bk} \frac{\ve_\bk-\xi_\bk}{\Delta \phi_\bk^*}  \sum_{s = \pm} \left( \bh_\bk \cdot {\boldsymbol \sigma} + s h_{\bk} \sigma^0 \right)\sigma^y(\sigma^0+s\sigma^z). \nonumber
\end{align}
The normalization constant ensuring the unitarity of $\hat U_\bk$ now equals 
\be  
X_\bk = \sqrt{\frac1{16 h_\bk (h_\bk - \bh_{\bk}\cdot \hat {\bf z})} \left( 1 + \frac{\xi_\bk}{\ve_\bk}  \right)}.
\ee

To derive the gap equation, we start again with the definition~\eqref{MainSC}. We find now for the anomalous average $\langle\hat{c}_{-\bk\mu}\hat{c}_{\bk\lambda}\rangle = - i {\cal F}_\bk{\sigma}^y_{\lambda \mu}$, where
\be \label{Eq:Fsinglet} 
{\cal F}_\bk = \frac{\Delta \phi_\bk}{4 \ve_\bk} \sum_{s = \pm}\tanh\left(\frac{E_{\bk,s}}{2T}\right).
\ee

Pairing potential in the spin-singlet channel can be decomposed as~\cite{MineevSamokhinbook} 
\be\label{Interactionnsinglet}
\begin{aligned}
V_{\alpha\beta,\lambda\mu}(\bk,\bp)=\frac{1}{2}(i\sigma^y)_{\alpha \beta} (i\sigma^y)^\dagger_{\lambda \mu} \\ \times\sum_{\Gamma}V_\Gamma(k,p)\sum\limits_{i=1}^{d_\Gamma}\psi_i^\Gamma(\hat{\bk}) \psi_{i}^{\Gamma*}(\hat{\bp}),
\end{aligned}
\ee
where $\psi_i^\Gamma(\hat \bk)$ are the orthonormal basis functions of a given irreducible representation $\Gamma$ and $V_\Gamma(k,p)$ are given by Eq.~\eqref{VGammakkp}.

Assuming the strongest instability is in a given representation while neglecting all others, we can write for the order parameter 
\beg \label{Eq:geta}
\phi_\bk=\sum\limits_{i=1}^{d_\Gamma}\eta_i\psi_i(\hat{\bk}), \qquad \sum\limits_{i=1}^{d_\Gamma}|\eta_i|^2=1. 
\en
Plugging Eqs.~\eqref{Eq:Fsinglet}-\eqref{Eq:geta}  into Eq.~\eqref{MainSC}, we find a set of gap equations for each $i=1,\ldots,d_\Gamma$:
\be\label{AuxilarySinglet} 
\Delta \cdot \eta_i = v_\Gamma \sum_{\bk} \psi_i^*(\hat{\bk})\cdot {\cal F}_\bk,
\ee 
which after taking into account Eq.~(\ref{Eq:Fsinglet}) eventually leads us to the self-consistency equation
\begin{align} \label{Eq:gapswave} 
&\frac1{v_\Gamma} = \sum_{\bk,s=\pm} \frac{|\phi_\bk|^2}{4\ve_\bk}\tanh\left(\frac{E_{\bk,s}}{2T}\right).
\end{align}

The energy of the condensate can easily be found as 
\be
\begin{aligned}
{\cal E}_0 &= \Delta\sum_{\bk} \phi_\bk \cdot {\cal F}^*_\bk + \frac{1}2\sum_{\bk,s}\left(\xi_{\bk,s} - |E_{\bk,s}|\right)
\\
&= \frac{\Delta^2}{v_\Gamma} + \frac{1}2\sum_{\bk,s}\left(\xi_{\bk,s} - |E_{\bk,s}|\right),
\end{aligned}
\ee
which coincides with Eq.~\eqref{E0Familiar}. 

Setting $\Delta\to0$ in Eq.~\eqref{Eq:gapswave}, we find for the second-order transition temperature~\cite{FrigerietallPRL2004} 
\be  \label{Eq:Tcsinglet}
\ln \left(\frac{T_c}{T_{c0}}\right) =  \left\langle |\phi_\bk|^2 \cdot F\left( \frac{h_\bk}{2T_c}  \right)\right\rangle_{\hat{\bk}},
\ee 
where function $F(x)$ is defined in Eq.~\eqref{Tc}. 

Finally, using Eq.~\eqref{DefineM}, we find for the spin magnetization in the spin-singlet state: 
\be  \label{Eq:Msinglet}
{\bf M} = \frac{\mu_B}4 \sum_{\bk,s=\pm}s \hat g^T_\bk \hat \bh_\bk \tanh\left(\frac{E_{\bk,s}}{2T}  \right),
\ee 
where $\hat \bh_\bk = \bh_\bk/h_\bk$ and $\hat g_\bk$ is the Landé-factor matrix. The zero-field spin susceptibility follows from this expression easily, reproducing the result from Ref.~\cite{Samokhin2021}: 
\be  \label{Eq:chi0singlet}
\chi_{ij}^{(0)}  = \frac{\chi_n}4\left\langle \left[ \hat g^T_\bk \hat g_\bk \right]_{ij} Y(\bk,T) \right\rangle_{\hat \bk},
\ee 
where the Yoshida function $Y(\bk,T)$ is defined in Eq.~\eqref{Eq:Yoshida}, with $|\bd_\bk|^2$ replaced by $|\phi_\bk|^2$, and $\chi_n = 2\mu_B^2 \nu_F$ is the normal-state Pauli susceptibility.

In the simplest case $[\hat{g}_\bk]_{ij} = 2 \delta_{ij}$, which implies that $\bh_\bk = \mu_B \bH$, Eqs.~\eqref{Eq:Tcsinglet}-\eqref{Eq:chi0singlet} reproduce the well-known textbook expressions: 
\begin{align}
&\ln \left(\frac{T_c}{T_{c0}}\right) =  F\left( \frac{\mu_B H}{2T_c}  \right), \nonumber \\ &{\bf M} = \frac{\mu_B}2 \hat\bH \sum_{\bk,s=\pm}s \cdot \tanh\left(\frac{E_{\bk,s}}{2T}  \right),\\ 
&\chi_{ij}^{(0)}  = \chi_n\delta_{ij}\left\langle Y(\bk,T) \right\rangle_{\hat \bk}. \nonumber
\end{align}

We notice that Eqs.~\eqref{Eq:gapswave}-\eqref{Eq:chi0singlet} formally coincide with the analogous results for the unitary spin-triplet superconductivity in the case when $\bd_\bk \parallel \bh_\bk$ for all $\bk$, and with $|\bd_\bk|^2$ being replaced by $|\phi_\bk|^2$. The crystal symmetry-allowed basis functions and gap structures are, of course, different for spin-singlet and spin-triplet states.

\begin{widetext}

\section{Green's functions formalism \label{App:Gorkov}}
In this Appendix, we show how the problem of a spin-triplet superconductor in Zeeman magnetic field can be solved by the method of Gor'kov Green's functions. Not only this method allows us to reproduce all the results obtained in this paper, but it is also very powerful and convenient for a range of other applications.

First, we introduce the following four component spinor in pseudospin and Nambu spaces:
\beg\label{MySpinor}
\begin{aligned}
\check{\Psi}_\bk&=\left(\begin{matrix} 
\hat c_{\bk\up} \\ \hat c_{\bk\dn} \\ -\hat c_{-\bk\dn}\dg \\ \hat c_{-\bk\up}\dg
\end{matrix}
\right),
\end{aligned}
\en
and also $\check{\Psi}_\bk\dg=\left(\hat c_{\bk\up}\dg ~\hat c_{\bk\dn}\dg ~-\hat c_{-\bk\dn} ~\hat c_{-\bk\up}\right)$. We emphasize that this choice of a Nambu spinor differs from Eq.~\eqref{Eq:cNambu} as it is more custom in literature that operates with the superconducting Green's functions. The band part of the Hamiltonian, Eq.~\eqref{EQ1}, then reads
\beg\label{Eq1}
\hat{\cal H}_0=\sum\limits_{\bk\alpha\beta}\hat{c}_{\bk\alpha}\dg\left[\xi_{\bk}\delta_{\alpha\beta}-({\mathbf h}_\bk\cdot{\mbox{\boldmath $\sigma$}})_{\alpha\beta}\right] \hat{c}_{\bk\beta}\to \frac12 \sum\limits_{\bk}\sum\limits_{\mathrm{ab}}\check{\Psi}_{\bk\mathrm{a}}\dg\left[\xi_\bk({\tau}_3\otimes{\sigma}_0)-{\tau}_0\otimes({\mathbf h}_\bk\cdot{\mbox{\boldmath ${\sigma}$}})\right]_{ab}\check{\Psi}_{\bk\mathrm{b}}.
\en
Here ${\tau}_i$ and ${\sigma}_j$ are Pauli matrices acting in Nambu and spin subspaces correspondingly. Similarly, the pairing term~\eqref{HDelta} equals
\beg\label{HpairFin}
\hat{\cal H}_\Delta=-\frac12\sum\limits_{\bk}\sum\limits_{\mathrm{ab}}\check{\Psi}_{\bk\mathrm{a}}\dg\left[{\tau}_1\otimes\left({\mathbf a}_\bk\cdot{\mbox{\boldmath ${\sigma}$}}\right)+{\tau}_2\otimes\left({\mathbf b}_\bk\cdot{\mbox{\boldmath ${\sigma}$}}\right)\right]_{\mathrm{ab}}\check{\Psi}_{\bk\mathrm{b}},
\en
where we have parametrized vector ${\mathbf d}_\bk={\mathbf a}_\bk-i{\mathbf b}_\bk$, with real ${\mathbf a}_\bk$ and ${\mathbf b}_\bk$, and included the pairing amplitude $\Delta$ into the definition of this vector for brevity. Given these expressions, the corresponding mean-field action for our problem is  
\beg\label{Saction}
{S}=\frac12\int\limits_{-\infty}^\infty dt\sum\limits_{\bk}\check{\Psi}\dg\left[i({\tau}_0\otimes{\sigma}_0)\frac{\partial}{\partial t}-\xi_\bk({\tau}_3\otimes{\sigma}_0)
+{\tau}_0\otimes({\mathbf h}_\bk\cdot{\mbox{\boldmath ${\sigma}$}})
+{\tau}_1\otimes\left({\mathbf a}_\bk\cdot{\mbox{\boldmath ${\sigma}$}}\right)+{\tau}_2\otimes\left({\mathbf b}_\bk\cdot{\mbox{\boldmath ${\sigma}$}}\right)\right]\check\Psi.
\en
Here, the summation over the spinor components has been assumed. Expression in the brackets in (\ref{Saction}) determines the inverse of the single-particle propagator:
\beg\label{DefProp}
\check{G}(1,2)=-i\left\langle\hat{T}_t\left\{\check\Psi(1)\otimes\check\Psi\dg(2)\right\}\right\rangle.
\en
The inverse of the retarded propagator in frequency-momentum space is given by 
\beg\label{InvG}{
\check{G}^{-1}(\bk,\omega)=(\omega+i0)({\tau}_0\otimes{\sigma}_0)-\xi_\bk({\tau}_3\otimes{\sigma}_0)
+{\tau}_0\otimes({\mathbf h}_\bk\cdot{\mbox{\boldmath ${\sigma}$}})
+{\tau}_1\otimes\left({\mathbf a}_\bk\cdot{\mbox{\boldmath ${\sigma}$}}\right)+{\tau}_2\otimes\left({\mathbf b}_\bk\cdot{\mbox{\boldmath ${\sigma}$}}\right).}
\en
 The single-particle spectrum can be found by setting the determinant of $\check{G}^{-1}(\bk,\omega)$ to zero. It is straightforward to show that
\beg\label{detGinv}
\textrm{det}\left\{\check{G}^{-1}(\bk,\omega)\right\}=(\omega^2-E_{\bk,+}^2)(\omega^2-E_{\bk,-}^2),
\en 
where $E_{\bk,\pm}$ satisfy the following equations:
\beg\label{Ebppm}
\begin{aligned}
&E_{\bk,+}^2+E_{\bk,-}^2=2\left(\xi_\bk^2+h_\bk^2+|{\mathbf d}_\bk|^2\right), \\ 
&E_{\bk,+}^2E_{\bk,-}^2=\left(\xi_\bk^2+h_\bk^2+|{\mathbf d}_\bk|^2\right)^2-4[({\mathbf a}_\bk\cdot{\mathbf h}_\bk)^2+({\mathbf b}_\bk\cdot {\mathbf h}_\bk)^2]+\left(\bq_\bk-2\xi_\bk{\mathbf h}_\bk\right)^2.
\end{aligned}
\en
Note that vector $\bq_\bk$ here absorbs the factor $\Delta^2$ and is defined according to
\beg\label{qvector}
\bq_\bk=i({\mathbf d}_\bk\times{\mathbf d}_\bk^*)=-2({\mathbf a}_\bk\times{\mathbf b}_\bk).
\en
Therefore, we have four solutions which are derived from
\beg\label{EpmFin}
{E_{\bk,\pm}^2=\xi_\bk^2+h_\bk^2+|{\mathbf d}_\bk|^2\pm\sqrt{4({\mathbf a}_\bk\cdot{\mathbf h}_\bk)^2+4({\mathbf b}_\bk\cdot{\mathbf h}_\bk)^2+\left(\bq_\bk-2\xi_\bk{\mathbf h}_\bk\right)^2}},
\en
which, of course, coincides with Eq.~\eqref{Eks} of the main text.

\subsection{Normal components of the correlation function}
For the spin-diagonal components of the normal correlation function
\beg\label{Gab}
{\cal G}_{\alpha\beta}(\bk;t,t')=-i\left\langle\hat{T}_t\left\{\hat c_{\bk\alpha}(t) \hat c_{\bk\beta}\dg(t')\right\}\right\rangle,
\en
which are determined by the top left $2\times 2$ block in \eqref{DefProp}, i.e. ${\cal G}_{\up\up}=[\check{G}]_{11}$,  ${\cal G}_{\up\dn}=[\check{G}]_{12}$, ${\cal G}_{\dn\up}=[\check{G}]_{21}$ and ${\cal G}_{\dn\dn}=[\check{G}]_{22}$.
We find
\beg\label{calGaa4M}
\begin{aligned}
{\cal G}_{\up\up}(\bk,\omega)&=\frac{(\omega+\xi-h_z)[\omega^2-(\xi+h_z)^2-|{\mathbf d}|^2-q_z]}{(\omega^2-E_{+}^2)(\omega^2-E_{-}^2)}-\frac{(h_x+ih_y)(d_x^*-id_y^*)d_z+(h_x-ih_y)(d_x+id_y)d_z^*}{(\omega^2-E_{+}^2)(\omega^2-E_{-}^2)}
\\&-\frac{(h_x^2+h_y^2)(\omega-\xi-h_z)+2h_z|{d}_z|^2}{(\omega^2-E_{+}^2)(\omega^2-E_{-}^2)}, \\
{\cal G}_{\dn\dn}(\bk,\omega)&=\frac{(\omega+\xi+h_z)[\omega^2-(\xi-h_z)^2-|{\mathbf d}|^2+q_z]}{(\omega^2-E_{+}^2)(\omega^2-E_{-}^2)}+
\frac{(h_x-ih_y)(d_x^*+id_y^*)d_z+(h_x+ih_y)(d_x-id_y)d_z^*}{(\omega^2-E_{+}^2)(\omega^2-E_{-}^2)}\\&-
\frac{(h_x^2+h_y^2)(\omega-\xi+h_z)-2h_z|{d}_z|^2}{(\omega^2-E_{+}^2)(\omega^2-E_{-}^2)}.
\end{aligned}
\en
Here, we have omitted the momentum subscripts for brevity. 
Similarly, for the off-diagonal components of $\hat{\cal G}(\bk,\omega)$, we obtain:
\beg\label{calGab4M}
\begin{aligned}
{\cal G}_{\up\dn}(\bk,\omega)&=-\frac{(h_x-ih_y)[(\omega+\xi)^2-h^2-|d_z|^2]}{(\omega^2-E_{+}^2)(\omega^2-E_{-}^2)}-\frac{(h_x+ih_y)[|d_x|^2-|d_y|^2-i(d_xd_y^*+d_x^*d_y)]}{(\omega^2-E_{+}^2)(\omega^2-E_{-}^2)} \\ & -\frac{(\omega+\xi)(q_x-iq_y)+h_z[(d_x^*-id_y^*)d_z+(d_x-id_y)d_z^*]}{(\omega^2-E_{+}^2)(\omega^2-E_{-}^2)}, \\
{\cal G}_{\dn\up}(\bk,\omega)&=-\frac{(h_x+ih_y)[(\omega+\xi)^2-h^2-|d_z|^2]}{(\omega^2-E_{+}^2)(\omega^2-E_{-}^2)}-\frac{(h_x-ih_y)[|d_x|^2-|d_y|^2+i(d_x^*d_y+d_xd_y^*)]}{(\omega^2-E_{+}^2)(\omega^2-E_{-}^2)} \\ & -\frac{(\omega+\xi)(q_x+iq_y)+h_z[(d_x^*+id_y^*)d_z+(d_x+id_y)d_z^*]}{(\omega^2-E_{+}^2)(\omega^2-E_{-}^2)}.
\end{aligned}
\en
Using these expressions, spin magnetization can be easily computed by going to the Matsubara frequency representation $\omega\to i\omega_n$ as
\beg\label{StatMag}
{\mathbf M}=-\frac{\mu_B}2T\sum\limits_{i\omega_n}\sum\limits_{\bk}\mathrm{Tr}\left[\hat g_\bk^T{\mbox{\boldmath $\sigma$}}\hat{\cal G}(\bk;i\omega_n)\right]e^{i\omega_n0+}.
\en
Evaluating summation over Matsubara frequencies, we end up with Eq.~\eqref{MExplicit} in the main text.

\subsection{Anomalous (Gor'kov) components of the correlation function}
As it follows from the general definition \eqref{Gab}, the anomalous (Gor'kov) correlation functions will be given by the off-diagonal block of $\check{G}$. Specifically, using the standard definition ${\cal F}_{\alpha\beta}(\bk;t-t')=i\langle\hat{T}_t\{c_{\bk\alpha}(t) c_{-\bk\beta}(t')\}\rangle$ we have
\beg\label{Fabexact}
\begin{aligned}
&{\cal F}_{\dn\up}(\bk,\omega)=[\check{G}(\bk,\omega)]_{24}, \quad {\cal F}_{\up\up}(\bk,\omega)=[\check{G}(\bk,\omega)]_{14}, \\
&{\cal F}_{\up\dn}(\bk,\omega)=-[\check{G}(\bk,\omega)]_{13}, \quad {\cal F}_{\dn\dn}(\bk,\omega)=-[\check{G}(\bk,\omega)]_{23}.
\end{aligned}
\en

We start with the spin off-diagonal  components:
\beg\label{calFoff4M}
\begin{aligned}
[\check{G}(\bk,\omega)]_{13}&=-\frac{(h_x+ih_y)(d_x-id_y)(\omega-\xi-h_z)}{(\omega^2-E_{+}^2)(\omega^2-E_{-}^2)}-\frac{(h_x-ih_y)(d_x+id_y)(\omega+\xi-h_z)}{(\omega^2-E_{+}^2)(\omega^2-E_{-}^2)}\\&+
\frac{d_z[(\omega-h_z)^2-\xi^2-|{\mathbf d}|^2-h_\perp^2]}{(\omega^2-E_{+}^2)(\omega^2-E_{-}^2)}+\frac{i(q_yd_x-q_xd_y)}
{(\omega^2-E_{+}^2)(\omega^2-E_{-}^2)}, \\
[\check{G}(\bk,\omega)]_{24}&=-\frac{(h_x+ih_y)(d_x-id_y)(\omega+\xi+h_z)}{(\omega^2-E_{+}^2)(\omega^2-E_{-}^2)}-\frac{(h_x-ih_y)(d_x+id_y)(\omega-\xi+h_z)}{(\omega^2-E_{+}^2)(\omega^2-E_{-}^2)}\\& -
\frac{d_z[(\omega+h_z)^2-\xi^2-|{\mathbf d}|^2-h_\perp^2]}{(\omega^2-E_{+}^2)(\omega^2-E_{-}^2)}-\frac{i(q_yd_x-q_xd_y)}
{(\omega^2-E_{+}^2)(\omega^2-E_{-}^2)},
\end{aligned}
\en
where $h_\perp=\sqrt{h_x^2+h_y^2}$.
Note that the signs in front of the terms proportional to $d_z$ are opposite, as it should be. Similarly,
\beg\label{overcalFoff4M}
\begin{aligned}
[\check{G}(\bk,\omega)]_{31}&=
-\frac{(h_x-ih_y)(d_x^*+id_y^*)(\omega-\xi-h_z)}{(\omega^2-E_{+}^2)(\omega^2-E_{-}^2)}
-\frac{(h_x+ih_y)(d_x^*-id_y^*)(\omega+\xi-h_z)}{(\omega^2-E_{+}^2)(\omega^2-E_{-}^2)}\\&+
\frac{d_z^*[(\omega-h_z)^2-\xi^2-|{\mathbf d}|^2-h_\perp^2]}{(\omega^2-E_{+}^2)(\omega^2-E_{-}^2)}-\frac{i(q_yd_x^*-q_xd_y^*)}
{(\omega^2-E_{+}^2)(\omega^2-E_{-}^2)}, \\
[\check{G}(\bk,\omega)]_{42}&=-\frac{(h_x-ih_y)(d_x^*+id_y^*)(\omega+\xi+h_z)}{(\omega^2-E_{+}^2)(\omega^2-E_{-}^2)}-\frac{(h_x+ih_y)(d_x^*-id_y^*)(\omega-\xi+h_z)}{(\omega^2-E_{+}^2)(\omega^2-E_{-}^2)}\\& -
\frac{d_z^*[(\omega+h_z)^2-\xi^2-|{\mathbf d}|^2-h_\perp^2]}{(\omega^2-E_{+}^2)(\omega^2-E_{-}^2)}+\frac{i(q_yd_x^*-q_xd_y^*)}
{(\omega^2-E_{+}^2)(\omega^2-E_{-}^2)}.
\end{aligned}
\en
We see that $[\check{G}(\bk,\omega)]_{31}^*=[\check{G}(\bk,\omega)]_{13}$ and $[\check{G}(\bk,\omega)]_{42}^*=[\check{G}(\bk,\omega)]_{24}$, which could have been expected. 

Next, we evaluate the spin-diagonal components in the upper block. We find:
\beg\label{calFdia4M}
\begin{aligned}
[\check{G}(\bk,\omega)]_{14}&=\frac{(h_x-ih_y)[(h_x-ih_y)(d_x+id_y)+2d_z(\xi+h_z)]}{(\omega^2-E_{+}^2)(\omega^2-E_{-}^2)}\\&+\frac{(d_x-id_y)[\omega^2-(\xi+h_z)^2-|{\mathbf d}|^2-q_z]}{(\omega^2-E_{+}^2)(\omega^2-E_{-}^2)}+\frac{(q_x-iq_y)d_z}
{(\omega^2-E_{+}^2)(\omega^2-E_{-}^2)}, \\
[\check{G}(\bk,\omega)]_{23}&=\frac{(h_x+ih_y)[(h_x+ih_y)(d_x-id_y)-2d_z(\xi-h_z)]}{(\omega^2-E_{+}^2)(\omega^2-E_{-}^2)}\\&+\frac{(d_x+id_y)[\omega^2-(\xi-h_z)^2-|{\mathbf d}|^2+q_z]}{(\omega^2-E_{+}^2)(\omega^2-E_{-}^2)}-\frac{(q_x+iq_y)d_z}
{(\omega^2-E_{+}^2)(\omega^2-E_{-}^2)}.
\end{aligned}
\en
Similarly, for the spin-diagonal components of anomalous functions from the lower block we have:
\beg\label{overcalFdia4M}
\begin{aligned}
[\check{G}(\bk,\omega)]_{41}&=\frac{(h_x+ih_y)[(h_x+ih_y)(d_x^*-id_y^*)+2d_z^*(\xi+h_z)]}{(\omega^2-E_{+}^2)(\omega^2-E_{-}^2)}\\&+\frac{(d_x^*+id_y^*)[\omega^2-(\xi+h_z)^2-|{\mathbf d}|^2-q_z]}{(\omega^2-E_{+}^2)(\omega^2-E_{-}^2)}
+\frac{(q_x+iq_y)d_z^*}{(\omega^2-E_{+}^2)(\omega^2-E_{-}^2)}, \\
[\check{G}(\bk,\omega)]_{32}&=\frac{(h_x-ih_y)[(h_x-ih_y)(d_x^*+id_y^*)-2d_z^*(\xi-h_z)]}{(\omega^2-E_{+}^2)(\omega^2-E_{-}^2)}\\&+\frac{(d_x^*-id_y^*)[\omega^2-(\xi-h_z)^2-|{\mathbf d}|^2+q_z]}{(\omega^2-E_{+}^2)(\omega^2-E_{-}^2)}
-\frac{(q_x-iq_y)d_z^*}
{(\omega^2-E_{+}^2)(\omega^2-E_{-}^2)}.
\end{aligned}
\en
The gap equations~\eqref{Eq:gapeqn1}-\eqref{SelfEq} can then be readily derived from the self-consistency condition
\beg\label{MainSCFin}
\Delta(T)=-\frac{V_{\Gamma}}{2}T\sum\limits_{i\omega_n}\sum_{\bk}\left({\mathbf d}^{*}_\bk\cdot{\mathbf g}_{\mu\lambda}\dg\right){\cal F}_{\lambda\mu}(\bk,i\omega_n)e^{i\omega_n0+}.
\en

\end{widetext}

\end{appendix}

\bibliography{Zeeman}

\end{document}